\documentclass[twocol]{ametsoc}
\makeatletter
\newcommand*{\rom}[1]{\expandafter\@slowromancap\romannumeral #1@}
\makeatother

\usepackage{amsmath,amsfonts,amssymb,bm}
\usepackage{mathptmx}
\usepackage{newtxtext}
\usepackage{newtxmath}

\bibpunct{(}{)}{;}{a}{}{,}

\title{The Transient Responses of An Axisymmetric Tropical Cyclone to Instantaneous Surface Roughening and Drying. Part \rom{1}: Numerical Experiments}

\authors{Jie Chen\correspondingauthor{Jie Chen, 
     Department of Earth, Atmospheric and Planetary Sciences, Purdue University, 550 Stadium Mall Dr., West lafayette, IN, 47907.}
 and Daniel R. Chavas}

\affiliation{Purdue University, West Lafayette, Indiana, USA}

\abstract{Inland tropical cyclone (TC) impacts due to high winds and rainfall-induced flooding depend strongly on the evolution of the wind field and precipitation distribution after landfall. However, research has yet to test the detailed response of a mature TC and its hazards to changes in surface forcing in idealized settings. This work tests the transient response of an idealized hurricane to instantaneous transitions in two key surface properties associated with landfall: surface roughening and drying. Simplified axisymmetric experiments are performed in CM1 where surface drag coefficient and evaporative fraction are each systematically modified beneath a mature hurricane. Surface drying stabilizes the eyewall and consequently weakens the overturning circulation, thereby reducing inward angular momentum transport that slowly decays the wind field only within the inner-core. In contrast, surface roughening initially ($\sim$12 hours) rapidly weakens the entire low-level wind field and enhances the overturning circulation dynamically despite the concurrent thermodynamic stabilization of the eyewall; thereafter the storm gradually decays similar to drying. As a result, total precipitation temporarily increases with roughening but uniformly decreases with drying. Storm size decreases monotonically and rapidly with surface roughening, while the radius of maximum wind can increase with moderate surface drying. Overall, this work provides a mechanistic foundation for understanding the inland evolution of real storms in nature.}

\begin{document}

\maketitle

\section{Introduction}
While the majority of existing tropical cyclone (TC) research has focused on its coastal impacts, TCs also induce significant inland hazards, especially damaging wind and rainfall-induced flooding. Economic loss can be significant due to agricultural impacts, power outage, and infrastructure damages associated with strong winds \citep{Berg2009,Cangialosi2018} and inland freshwater flooding \citep{Villarini2010,Villarini2014}. From 1970 to 1999, 63\% of all TC related deaths were inland deaths \citep{Rappaport2000}. Among the 2,325 individual deaths in the United States from 1963 to 2012, rainfall-induced flood deaths occurred more than any other hazard, whereas 5-10\% of the deaths were caused by nontornadic winds \citep{Rappaport2014}. For example, the slow-moving Hurricanes Harvey (2017) and Florence (2018) induced catastrophic inland flooding over Texas and the Carolinas, respectively \citep{Blake+Zelinsky2018}. Moreover, rainfall production is intimately linked to the TC low-level wind field \citep{Lu2018}. Therefore, credible estimates of inland damage risk depend on an understanding of how the TC wind and precipitation fields evolve after landfall. This is increasingly important given that storms are moving more slowly on average \citep{Kossin2018}, a trend that is expected to continue in a future climate \citep{Emanuel2017}. However, the response of a mature TC and associated hazards to landfall and its underlying physical processes has not been systematically tested in an idealized setting and remains poorly understood. This lack of knowledge greatly inhibits our ability to predict inland hazards and estimate hazard risk in both operations and long-term risk assessment.

Research has analyzed TC landfall using historical data, climate models, and real-world landfall simulations \citep{Powell1987,Powell1991,Davis2008,Jian+Wu2008,Kruk2010,Lin2010,Murakami2016}, and empirical models exist to predict the post-landfall decay of storm intensity \citep{Tuleya1994, Kaplan+DeMaria1995}. Though these studies are essential for analyzing individual case studies and for directly estimating real-world risk, the complexity of the real-world makes it difficult for such simulations to provide a fundamental understanding of the response of a storm to changes in surface properties associated with landfall. A limited number of idealized studies or sensitivity tests exist, though they focus primarily on the evolution of storm intensity \citep{Tuleya+Kurihara1975,Tuleya+Kurihara1978, Kishtawal2012}, or the overall asymmetric wind distribution \citep{Wong+Chan2007}, and they are ripe for updating given advances in theory or numerical modeling since their publication. A few studies have applied theoretical models to examine the post-landfall response of the boundary layer \citep{Powell1982,Kepert2012} and the precipitation distribution associated with real-world storms \citep{Kepert2013,Lu2018}. However, systematic, idealized experiments testing the response of the tropical cyclone wind and precipitation distributions to landfall have yet to be performed.

Meanwhile, decades of research have advanced our understanding of tropical cyclones over an ocean surface that may also be useful for understanding the post-landfall evolution \citep{Malkus1960,Ooyama1964,Ooyama1969,Anthes1974,Emanuel1986}. In particular, TC Potential Intensity theory \citep{Emanuel1986} provides a natural starting point for understanding the response to landfall. This theory idealizes a mature, steady-state tropical cyclone as a Carnot heat engine, in which entropy fluxes from the ocean surface are used to maintain the circulation against surface frictional dissipation. It has also been extended to predict time-dependent changes in storm intensity including the effects of ventilation of environmental air by wind shear \citep{Tang2010,Tang2012}. This theory predicts the maximum potential intensity that a storm may achieve in a given thermodynamic environment as:
\begin{equation}\label{EQvp}
    V_p=\sqrt{\frac{C_k}{C_d}\eta(\triangle {k})} 
\end{equation}
where
\begin{equation}\label{EQ_eta}
\eta=\frac{T_{SST}-T_{tpp}}{T_{tpp}}
\end{equation}
\begin{equation}\label{EQvp_k}
    \triangle{k}={C_p}(T_{SST}-T_a)+L_v(q^*(T_{SST})-q_a(T_a))
\end{equation}
$C_k$ and $C_d$ are bulk exchange coefficient for surface enthalpy and
momentum; $\triangle{k}$ is the difference between the saturation enthalpy of the ocean surface and the enthalpy of the overlying near-surface air; $T_{SST}$ is the sea surface temperature, $T_a$ is the temperature of air overlying the ocean surface; $T_{tpp}$ is the tropopause temperature; $L_v$ is the enthalpy of vaporization; $C_p$ is the specific heat capacity of air; $q^*$ is the saturation mixing ratio of the ocean surface at the local surface pressure; and $q_a$ is the mixing ratio of air overlying the ocean surface. Eq.\ref{EQvp} predicts that the equilibrium intensity will decrease with higher $C_d$ (i.e., a rougher surface) and smaller $\triangle{k}$ (i.e., a drier surface), both of which are associated with a transition from the ocean to the land surface. Indeed, storm intensity is well-known to decay rapidly after landfall in both observations and real-world numerical simulations \citep{Kaplan+DeMaria1995,Knutson2015}, though there are rare cases where storms maintain their intensity temporarily after the landfall, due perhaps to locally enhanced surface heat fluxes or baroclinic enhancement \citep{Emanuel2008,Evans2011}. Moreoever, the existence of shallow land surface water can reduce the landfall decay rate \citep{Shen2001}. Preceding formal Potential Intensity theory, \citep{Ooyama1964} first demonstrated the essential role of surface enthalpy fluxes to a mature TC in a numerical model. Moreover, \cite{Tuleya+Kurihara1978} performed idealized tests of the sensitivity of TC intensity to surface roughness and surface evaporation, finding that the suppression of evaporation is the dominant factor for the decay of TC intensity after landfall. However, little follow-up work has been performed to test these responses in the context of modern potential intensity theory or to extend the study to the complete low-level wind field.

Additionally, recent work has advanced our understanding of the complete wind field over ocean \citep{Colon1963,Eliassen1971,Eliassen+Lystad1977,Merrill1984,Holland2010,Chavas2016}. For tropical cyclones over the ocean, a theoretical model for the complete wind field now exists (\cite{Chavas2015}, hereafter C15) that combines models for the inner, convecting region wind field \citep{Emanuel+Rotunno2011} and outer, non-convecting region wind field \citep{Emanuel2004}. This model can capture the first-order structure and variability of the complete low-level wind field over ocean \citep{Chavas+Lin2016} and can also explain equilibrium wind field structure across both moist and dry surfaces \citep{Cronin+Chavas2019}. However, this model has yet to be tested for the transient response after landfall. 

In terms of rainfall, extensive observational and numerical studies have been performed to understand key aspects of the inland rainfall evolution. Asymmetric rainfall characteristics, including mean rain rate, distribution, and total precipitation, are related to landfalling TC intensity on average \citep{Lonfat2007,Kimball2008,Liu2018}, though peak rainfall is not correlated with TC intensity \citep{Deng2017}. Rainfall asymmetry is often forced by environmental vertical wind shear, nonuniform surface characteristics, and mesoscale convective
activity rather than TC intensity \citep{Rogers2003,Chan2004,Chen2006,Hsu+Kuo2013,Li+Duan2013,Li2014,Meng+Wang2016}. More recently, a physics-based TC rainfall model has been developed for risk analysis to model real-world rainfall risk \citep{Zhu2017,Emanuel2017}. This model quantifies contributions to rainfall from frictional convergence, topographic uplift, vortex stretching, and baroclinic forcing \citep{Lu2018}. However, research has yet to test the transient precipitation response to landfall and its underlying physical mechanisms in an idealized setting.

This work seeks to fill the above knowledge gaps by systematically testing the response of a mature, axisymmetric hurricane to instantaneous surface roughening and drying in an idealized setting.  Specifically, we seek to answer the following research questions in this work:
\begin{enumerate}
    \item In addition to intensity, how does the structure of the wind and precipitation fields respond to surface roughening or drying?
    \item What are the time-scales of these responses, and how do they differ between each surface forcing?
    \item What is the radial structure of the response, and how does it differ between each surface forcing? 
    \item Can we understand the physical mechanisms that govern these responses?
\end{enumerate}

To answer the above questions, we perform and analyze two sets of experiments in which the surface beneath an initially quasi-steady hurricane is instantaneously dried or roughened over a range of magnitudes (Section \ref{methods}). We first characterize the basic transient responses of TC intensity, size, and precipitation across all experiments (Section \ref{response1}). We then focus on one representative experiment from each set for an in-depth, comparative analysis of the radial structure of the response (Section \ref{response2}). For these two cases, we further isolate physical processes governing the responses of the low-level wind field via absolute angular momentum budget analysis and of the precipitation responses via a simple dynamic-thermodynamic decomposition (Section \ref{mechanism}). Finally, we summarize the major findings of this work and discuss limitations and avenues for future works (Section \ref{summary}).

\section{Methods}\label{methods}
\subsection{Overview}
As noted above, tropical cyclone landfall is associated with two basic surface forcings acting on a mature hurricane: surface roughening and surface drying. Here we seek to analyze the transient responses of the TC vortex and precipitation to these two forcings in idealized experiments in which each forcing is applied instantaneously over a range of magnitudes. We note that, for real-world landfall, inland surface roughness and wetness have significant spatiotemporal variations \citep{Cosby1984,Stull1988}. Surface drag coefficient can range from 0.002 to 0.3 base on different terrain types \citep{Holmes2001}. Representing such complexity requires sophisticated land-surface schemes and boundary layer schemes to parameterize soil, vegetation, and land use \citep{Zhang2017,Zhang2019,Davis2008,Nolan2009,Jin2010,Kishtawal2012} that translate to large spatiotemporal variation in surface roughness and wetness that will vary from storm to storm. These simulations provide essential insight for simulating the detailed evolution of real-world storms. Here, though, we seek a more general understanding of the most fundamental surface forcings associated with landfall in a simplified setting absent the pronounced spatiotemporal heterogeneity in surface properties found in the real world. Hence, we focus on experiments testing the TC transient response to surface drying and roughening in an axisymmetric geometry with a uniform environment and boundary forcing. Future work may add additional types of complexity to understand the effects that arise from asymmetries in the storm, surface, or environment. This idealized work serves as a rung in the hierarchy of models \citep{Held2005} that, in conjunction with more complex idealized experiments and real-world simulations, will improve both the basic understanding of landfall and the prediction of its inland hazards.

\subsection{Model description}
Numerical simulation experiments in axisymmetric geometry are performed using the Bryan Cloud Model (CM1v19.7). CM1 is suitable for use in a broad range of atmospheric science applications across scales, including hurricanes \citep{Chavas+Emanuel2014,Peng2018} and severe storms \citep{Sherburn+Parker2019,Trapp2019}. CM1 satisfies near-exact conservation of both mass and energy in a reversible saturated environment \citep{Bryan+Fritsch2002,Rotunno+Bryan2012}. The model solves the fully compressible equations of motion in height coordinates on an f-plane for flow velocities $(u, v, w)$, non-dimensional pressure $\pi$, potential temperature $\theta$, and the mixing ratios of water in vapor, liquid, and solid states $q_x$ on a fully staggered Arakawa C-type grid.

\subsection{Model set-up}
The basic axisymmetric, $f$-plane model set-up is similar to \cite{Chavas+Emanuel2014} (Table.\ref{geom}). The Coriolis parameter is set to $f=5\times 10^{-5} \; s^{-1}$. The outer wall is set to $L=3000\;km$ with 3-km radial grid spacing. A stretched grid is used in vertical with a constant grid spacing of 100-m below $z=3\;km$, smoothly-varying vertical grid spacing from 100-m at $z=3\;km$ to 500-m at $z=12\;km$, and then constant grid spacing of 500-m from $z=12\;km$ to $z=25\;km$. 
\begin{table}
\caption{Parameter values of the baseline and CTRL simulation.}
\label{geom}
\begin{tabular}{ |p{1.5cm} p{3.5cm} p{1.5cm}| }
\hline
Model& Name& Value \\
\hline
 $l_h$ &horizontal mixing length & 750-m\\
 $l_{inf}$& vertical mixing length&  100-m\\
$C_k \And C_d$ & Exchange Coef.of enthalpy and momentum &0.0015\\
$H_{domain}$&model height& 25-km \\
 $L_{domain}$&model radius&3000-km\\
\hline 
Environment& Name & Value\\
\hline 
$T_{sst}$&  sea surface temperature &300-K\\
$T_{tpp}$ &tropopause temperature   &200-K \\
$Q_{cool}$& radiative cooling& $1 K day^{-1}$\\
$f$&Coriolis Parameter& $5\times 10^{-5} s^{-1}$\\
\botline
\end{tabular}
\end{table}
In axisymmetric geometry, turbulent eddies cannot be resolved directly and thus are parameterized using a modified Smagorinsky-type scheme with distinct mixing lengths in the radial ($l_h=750\;m$, constant) and vertical directions (asymptotes to $l_{inf} = 100\;m$ at $z=\infty$). In CM1v19.7, the horizontal Rayleigh damper ($hrdamp=1$) is only applied to the vertical velocity at lateral boundaries, and not to the horizontal velocities, to minimize artificial sources of momentum. The upper-level Rayleigh damping ($irdamp=1$) is applied above $z=20\;km$, damping horizontal and vertical velocities towards the base state. The radiation scheme simply applies a constant cooling rate, $Q_{cool}$, to the potential temperature where the absolute temperature exceeds a threshold temperature, $T_{tpp}$, and is otherwise relaxed back to $T_{tpp}$:
\begin{equation}\label{Qcool}
 \frac{\partial \theta}{\partial t}=\bigg\{
\begin{array}{c c}	
     \frac{\theta(p,T_{tpp})-\theta(p,T)}{\tau_{strat}} & T<T_{tpp}\\
      -Q_{cool}& T>T_{tpp}
\end{array}
\end{equation}
We set $Q_{cool}=-1 \; K \; day^{-1}$ and $T_{tpp}=200 \; K$, which corresponds to an approximate tropopause temperature. The Newtonian relaxation time-scale is $\tau_{strat}=12\;h$. This simple approach neglects all water-radiation and temperature-radiation feedbacks. A similar set-up applied to absolute temperature is also used in \cite{Cronin+Chavas2019}. Surface latent heat fluxes, $F_{LH}$, and sensible heat fluxes, $F_{SH}$, are calculated from the bulk-aerodynamic formulae of surface mixing ratio fluxes $F_{qv}$ and potential temperature fluxes $F_{\theta}$ in CM1 (sfcphysics.F) as 
\begin{equation}\label{Eq_LHflux}
F_{LH}= \rho L_v F_{qv}
\end{equation}
\begin{equation}\label{Eq_SHflux}
F_{SH}= \rho C_p F_{\theta}
\end{equation}
\begin{equation}\label{Eq_qvflux}
F_{qv}= \epsilon s_{10}  C_q  \Delta q
\end{equation} 
\begin{equation}\label{Eq_thflux}
F_{\theta}= s_{10} C_h  \Delta \theta
\end{equation}
where $C_q$ and $C_h$ are the exchange coefficients for the surface moisture and sensible heat, respectively; $s_{10}$ is the 10-m wind speed; $\Delta q$ and $\Delta \theta$ are the moisture and heat disequilibrium between the 10-m layer and the sea surface, respectively; and $\epsilon$ is the surface evaporative fraction ($\epsilon=1$ over ocean). $C_q$ and $C_h$ are set equal to each other, and so both represent the single enthalpy exchange coefficient, $C_k$. Surface drying is modeled by decreasing $\epsilon$ to represent the transition from a wet to a drier surface, thus decreasing the surface latent heat fluxes $F_{LH}$.

Surface roughening is modeled by increasing the drag coefficient $C_d$, which modifies the surface roughness length $z_0$ and, in turn, the friction velocity $u^*$ for the surface log-layer in CM1 as
\begin{equation}\label{Eq_Cd_z0}
z_0= \frac{z}{e^{(\frac{\kappa}{\sqrt{C_d}}-1)}}
\end{equation}
\begin{equation}\label{Eq_Cd_ust}
u^*= max\left[\frac{\kappa s_1}{ln(\frac{z_a}{z_0}+1)}, 1.0^{-6}\right]
\end{equation} where $\kappa$ is the von K\'arm\'an constant; $z=10m$ is the reference height; $s_1$ is the total wind speed on the lowest model; and $z_a$ is approximately equal to the lowest model level height. On the lowest model level, $u^*$ determines the subgrid shear stress $\tau_{i,j}$ (e.g., $\tau_{2,3}={u^*}^2 v_1/s_1$, where $(i,j)=(2,3)$ indicates the $v-w$ shear stress), which is used to calculate the turbulent tendencies. For the azimuthal wind, the turbulent tendency is:
\begin{equation}\label{Eq_Cd_turb}
   T_v=\frac{1}{\rho}\lbrack\frac{\partial \tau_{12}}{\partial x}+\frac{\partial \tau_{22}}{\partial y}+\frac{\partial \tau_{23}}{\partial z} \rbrack
\end{equation}
in the azimuthal velocity tendency equation given by
\begin{equation}\label{vten_full}
   \frac{\partial v}{\partial t}=-u\frac{\partial v}{\partial r}-w\frac{\partial v}{\partial z}+T_v+D_v
\end{equation} where $D_v$ represents the diffusive tendency. Hence, instantaneous changes in the parameters $C_d$ and $\epsilon$ are applied to model surface roughening and drying, respectively. 

We note that  $\epsilon$ is specifically applied to the latent fluxes $F_{LH}$, as surface enthalpy fluxes for mature storms over ocean are dominated by latent heat \citep{Charney1964,Kuo1965,Ooyama1969,Cione2000,Guimond2011}. Sensible heat fluxes $F_{SH}$ can become important when latent heat fluxes are very strongly reduced ($\epsilon < 0.1$), in which case the sensible heating provides a significant fraction of the total surface enthalpy fluxes. This can be understood via Eq.\eqref{EQvp} and is discussed in more detail in Appendix A. This suggests that land surface properties may become important for weak inland storms over very dry surfaces, as is occasionally observed with real storms \citep{Evans2011, Kieu2015,Shen2001}. For our work here, applying $\epsilon$ to sensible heat fluxes in addition to latent heat fluxes does not materially change our results. A more proper accounting of surface sensible heat fluxes likely requires experiments using a coupled land-atmosphere model that represent soil heat capacity and thermal diffusivity \citep{Emanuel2008,Kishtawal2012}. Thus, the role of changes in sensible heat fluxes is left for future work.

Using the above setup, we define an initial sounding as the domain-mean state for the final ten days of a 100-day simulation that has reached statistical equilibrium. This approach is commonly used in idealized modeling studies of tropical convection \citep{Wing+Emanuel2014,Wing2016,Peng2018}. Given that we begin from a long-run simulation (described below), our results are not sensitive to this choice of the initial state, though, as was found in \cite{Chavas+Emanuel2014}.
\begin{figure}[t]
\centerline{\includegraphics[width=20pc]{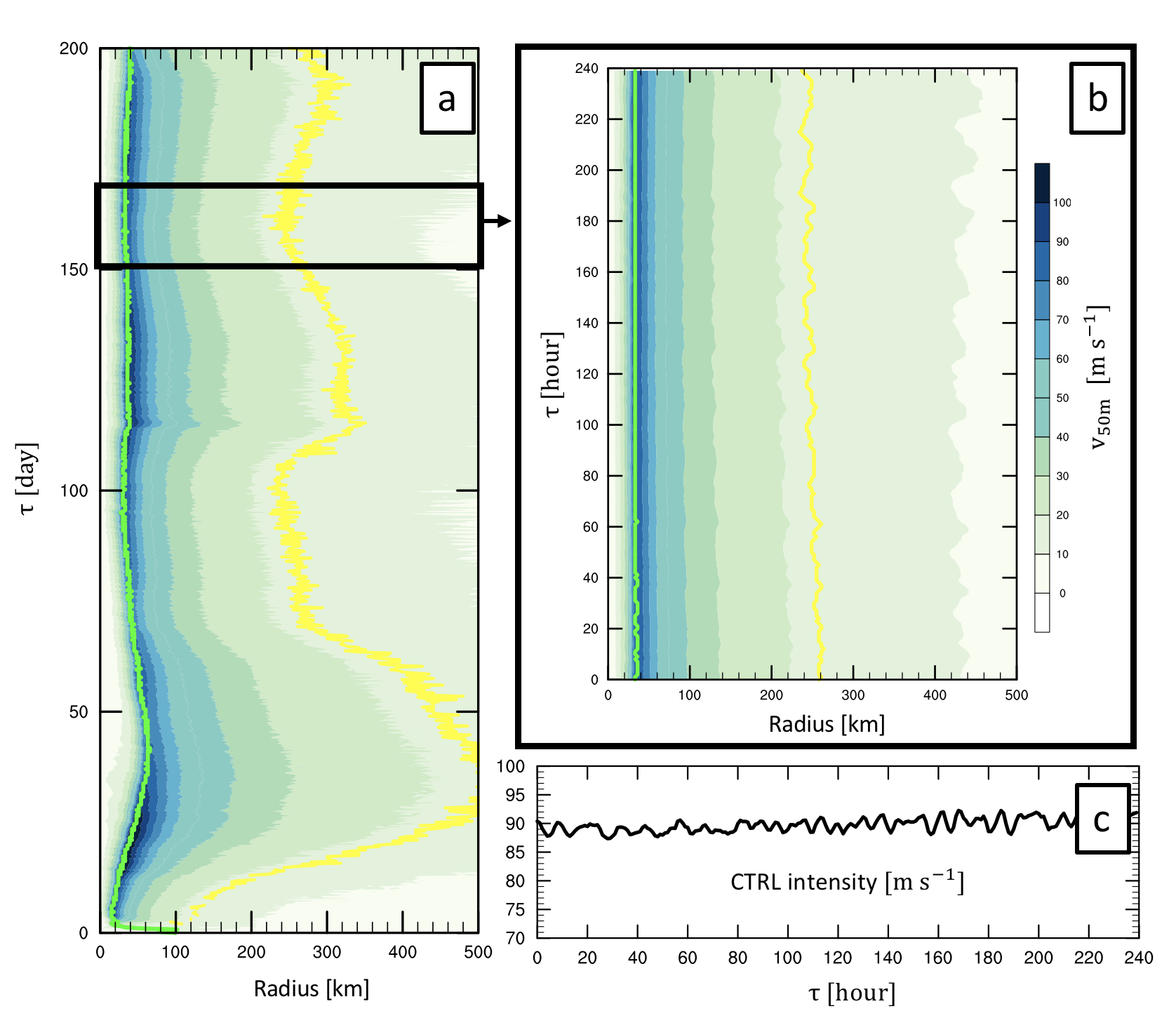}}
\caption{Hovm\"{o}ller diagram of the tangential wind field at the lowest model level ($50\;m$) for (a) the baseline 200-day ocean-surface simulation; (b) the ensemble-mean 10-day CTRL experiment generated from days 150-165 of the baseline simulation; (c) intensity ($v_m$) evolution for CTRL. Contours in Hovm\"{o}ller diagram denote $r_{max}$ (green) and $r_{34kt}$ (yellow).}\label{Hov}
\end{figure}
\subsection{Experimental design}
Before conducting surface roughening and drying experiments, we first create a baseline experiment that produces a tropical cyclone over an ocean-like surface where $C_d=0.0015$, $C_k=0.0015$, and $\epsilon=1$. We run this baseline experiment for 200 days to allow the storm to reach a statistical steady-state for at least 50 days (Fig.\ref{Hov}a). We then identify the most stable period during this long-term evolution, here selected as days 150-165; results are not sensitive to this choice of restart time. From this stable period, we define a 10-day Control experiment (CTRL) as the ensemble-mean of five 10-day segments of the baseline experiment as day150-160, day151-161, day152-162, day153-163 and day154-164 (Fig.\ref{Hov}b). We then perform ensembles of restart experiments with $C_d$ or $\epsilon$ instantaneously modified beneath the stable storm. $C_d$ is enhanced by a factor of 2, 4, 6, 8, and 10, while $\epsilon$ is reduced from 1 to 0.7, 0.5, 0.3, 0.25, and 0.1, representing various levels of surface roughening or drying, respectively. For each roughening or drying experiment, we also run five 10-day experiments restarting from each of the five CTRL ensemble segments. These five runs are then averaged into a 10-day ensemble representing the corresponding experiment. This ensemble approach reduces noise and increases the robustness of the responses. We use $\tau$ to denote the time since the start of a given experiment.

Note that $C_k$ may also vary in the real world, though there remain large uncertainties regarding $C_k$ and the ratio $\frac{C_k}{C_d}$ \citep{Bell2012}. A previous study suggests that the ratio of $\frac{C_k}{C_d}$ has less effect on the maximum wind speed in numerical simulations than in theoretical models, though the maximum wind speed is still proportional to $\frac{C_k}{C_d}$ \citep{Bryan2012}. Little is known about its inland variability nor how $C_k$ might scale with $C_d$ over land. For our work we simply hold $C_k$ constant, which translates to a decrease in the ratio of $\frac{C_k}{C_d}$ with roughening. Further discussion of treatment of the ratio of $\frac{C_k}{C_d}$ are provided in Section \ref{summary}. Our aim is to test the transient response specifically to enhanced $C_d$. The transient response of TC to surface roughening holding $\frac{C_k}{C_d}$ fixed will be the subject of future work.

\subsection{Analysis}
We first characterize the temporal evolution of the transient responses across all experiments by normalizing the evolution of storm intensity $v_m$ (maximum tangential wind speed at lowest model level), size $r_{max}$ and $r_{34kt}$ (radius of maximum tangential wind speed and $34kt$ tangential wind), and total precipitation $P$ (hourly total rainfall within r=150km) each by their time-dependent CTRL values.

We use the potential intensity equation (Eq.\ref{EQvp}) to define a theoretical prediction for the equilibrium response in maximum wind speed given by
\begin{equation}\label{predictedVm}
   \frac{{V_p}_{EXP}}{{V_p}_{CTRL}}=\frac{\sqrt{\frac{C_k}{(C_d)_{EXP}}\eta (C_p\triangle T +L_v(\triangle{q})_{EXP})}}{\sqrt{\frac{C_k}{C_d}\eta \triangle k}}
\end{equation} for comparison against the simulated response given by $\frac{{v_m}_{EXP}}{{v_m}_{CTRL}}$. $(C_d)_{EXP}$ is defined by its experimental value and $(\triangle{q})_{EXP}$ is defined by multiplying its CTRL value by $\epsilon$. The value of ${V_p}_{CTRL}$ is $69.93\;ms^{-1}$ with $\triangle T$ and $\triangle{q}$ defined from the environment (averaged over $r=1500-2500\;km$) and $66.37\;ms^{-1}$ defined at $r_{max}$. Both values are smaller than the actual CTRL $v_m$ on the lowest model level, 50-m ($90.4\;ms^{-1}$). Past work has shown that $v_m$ can exceed the potential intensity in nature and simulations \citep{Persing2003, Yang2007, Bryan+Rotunno2009a,Bryan+Rotunno2009b, Rousseau-Rizzi+Emanuel2019}. Moreover, the model's radial mixing length has a strong influence on the actual $v_m$ in axisymmetric simulations \citep{Bryan+Rotunno2009a, Chavas+Emanuel2014}. We do not address those complex issues here; instead, our focus is on the responses relative to CTRL. Notably, this normalized response approach conveniently yields very similar results when calculating $V_p$ from environmental data vs at the radius of maximum wind (Supplementary Table.1), with slightly larger disagreement for $0.1\epsilon$ experiment where sensible heat fluxes become increasingly important (Appendix and Fig.A1a and b). In our analyses below, we use $V_p$ calculated from the environment, which may be considered as a true environmental parameter.

We then seek to analyze the responses of the radial structure of the low-level (50-m) tangential wind field and precipitation field in greater depth using a representative experiment for each experiment type: $0.25\epsilon$ and $4C_d$. These two experiments yield a similar predicted potential intensity response of approximately $0.5$ (Eq.\ref{predictedVm}). Moreover, experimentally, they produce strong responses while also retaining a fully coherent vortex throughout the experiment (Section\ref{response1}). These two representative experiments are further used to understand the physical mechanisms underlying the responses. 

To understand the wind field responses, we perform budget analysis for the response in the absolute angular momentum (per unit mass), $M$, given by 
\begin{equation}\label{AAM}
  M=vr+\frac{1}{2}f r^2  
\end{equation} where $v$ is the tangential wind, $r$ is radius, and $f$ is the Coriolis parameter. $M$ is widely used for tropical cyclones since the radial structure of the tangential wind field is directly linked to the radial distribution of $M$ via Eq. \eqref{AAM}. Moreover, $M$ is theoretically convenient, as it increases monotonically with radius and is nearly conserved away from frictional boundaries in steady state \citep{Emanuel1986,Chavas2015,Emanuel+Rotunno2011,Peng2018}. Note that $M$ may temporarily decrease locally with radius \citep{Smith+Montgomery+Bui2018}, though such a state is inertially-unstable.

We define the absolute angular momentum response, $\Delta M$, as the difference between experiment and CTRL. To understand key processes governing the evolution of this response, we then define a budget equation for $\Delta M$. This begins from the traditional angular momentum budget, which is commonly used in past studies to understand the dynamics of storm circulation \citep{Tuleya+Kurihara1975}. In CM1, the budget equation for $M$ is formulated from the $v$-tendency equation as 
\begin{equation}\label{MbgtAlys}
    \frac{\partial M}{\partial t}=-u\frac{\partial M}{\partial r}-w\frac{\partial M}{\partial z}+rT_v+rD_v
\end{equation}
The first two terms on the RHS of the Eq.\ref{MbgtAlys} are the $M$ tendencies associated with radial and vertical advection; $rT_v$ is the tendency of $M$ due to turbulent stresses including surface drag; and $rD_v$ is the tendency of $M$ due to diffusion ($T_v$ and $D_v$ are applied directly to the $v$-tendency equation within CM1). Within the boundary layer, the direct loss of $M$ due to frictional dissipation is represented by negative values of $rT_v$. As noted above, subgrid stress terms on the model bottom boundary are modulated by $C_d$ via the frictional velocity $u^*$. Increasing $C_d$ enhances the surface sink of $M$ (for cyclonic flow) via the $rT_v$ term since $T_v$ is a function of the subgrid stress. Here the first two terms of Eq. \eqref{MbgtAlys} are calculated from model output of $v$-tendency, while the latter two are calculated from model output for $T_v$ and $D_v$. Budget analysis indicates that the diffusion tendency term $rD_v$ is not important and so will not be discussed further in this study. From Eq. \eqref{MbgtAlys}, we may then define a budget equation for $\Delta M$, given by
\begin{equation}\label{MbdgtDiff}
    \frac{\partial (\triangle{M})}{\partial t} = \triangle(\frac{\partial M}{\partial t})=\triangle(-u\frac{\partial M}{\partial r})+\triangle(-w\frac{\partial M}{\partial z})+\triangle(rT_v)
\end{equation}
Eq. \eqref{MbdgtDiff} can be applied directly to compare and contrast the underlying processes governing the angular momentum responses in each representative simulation. To minimize transient noise unrelated to the dynamics of spin down, for snapshot analyses we average our minute-by-minute model outputs every hour (1-hour average centered on given time). To understand the precipitation responses, we decompose the precipitation response into dynamic and thermodynamic contributions following a simplified decomposition, which is detailed in Section \ref{mechanism}\ref{precipitation}.

Finally, to understand both $M$ and precipitation responses, we also quantify the dynamical responses of the secondary circulation and the thermodynamic responses in surface mixing ratio fluxes and equivalent potential temperature. The secondary circulation is commonly quantified using the mass streamfunction, $\psi$ \citep[e.g.][]{Willoughby1979}. In axisymmetric geometry, the streamfunction is defined as
\begin{equation}\label{MassSF}
    \frac{\partial \psi}{\partial z}=-r\rho u, \;\;\;\; \frac{\partial \psi}{\partial r}=r\rho w
\end{equation}
$\psi$ can be approximated by integrating the first equation vertically upwards or the second equation radially outwards, e.g.
\begin{equation}
   \psi= \int_0^r r\rho w\, dr
\end{equation}
where $\psi=0$ at inner boundary $r=0$ km. Because our vertical grid is stretched, which introduces small integration errors, we integrate the radial equation to calculate $\psi(r,z)$ from model output as
\begin{equation}\label{MassSF3}
   \psi(r_{i+\frac{1}{2}},z)=\sum_{i=1}^{N} r_{i,z}\rho_{i,z} w_{i,z}\ \delta r
\end{equation} where $\rho_{i,z}$,$w_{i,z}$, and $r_{i,z}$ are the density, vertical velocity and radius of the $i$th radial grid box; $\delta r = 3 \; km$ is the radial grid resolution; and $i=1$ corresponds to $\frac{\delta r}{2}=1.5\;km$. We define the secondary circulation response, $\triangle{\psi}$, as the difference in mass stream function between experiments and CTRL. Note that $\psi$ is most accurately calculated by solving a Poisson equation for the entire flow field, though this is much more complex particularly given our vertically-stretched grid. Our simpler approach is common in practice \citep{Cook2004,Liu2007}, including in the NCAR Command Language \citep{NCL}.
\begin{figure}[t]
\centerline{\includegraphics[width=18pc]{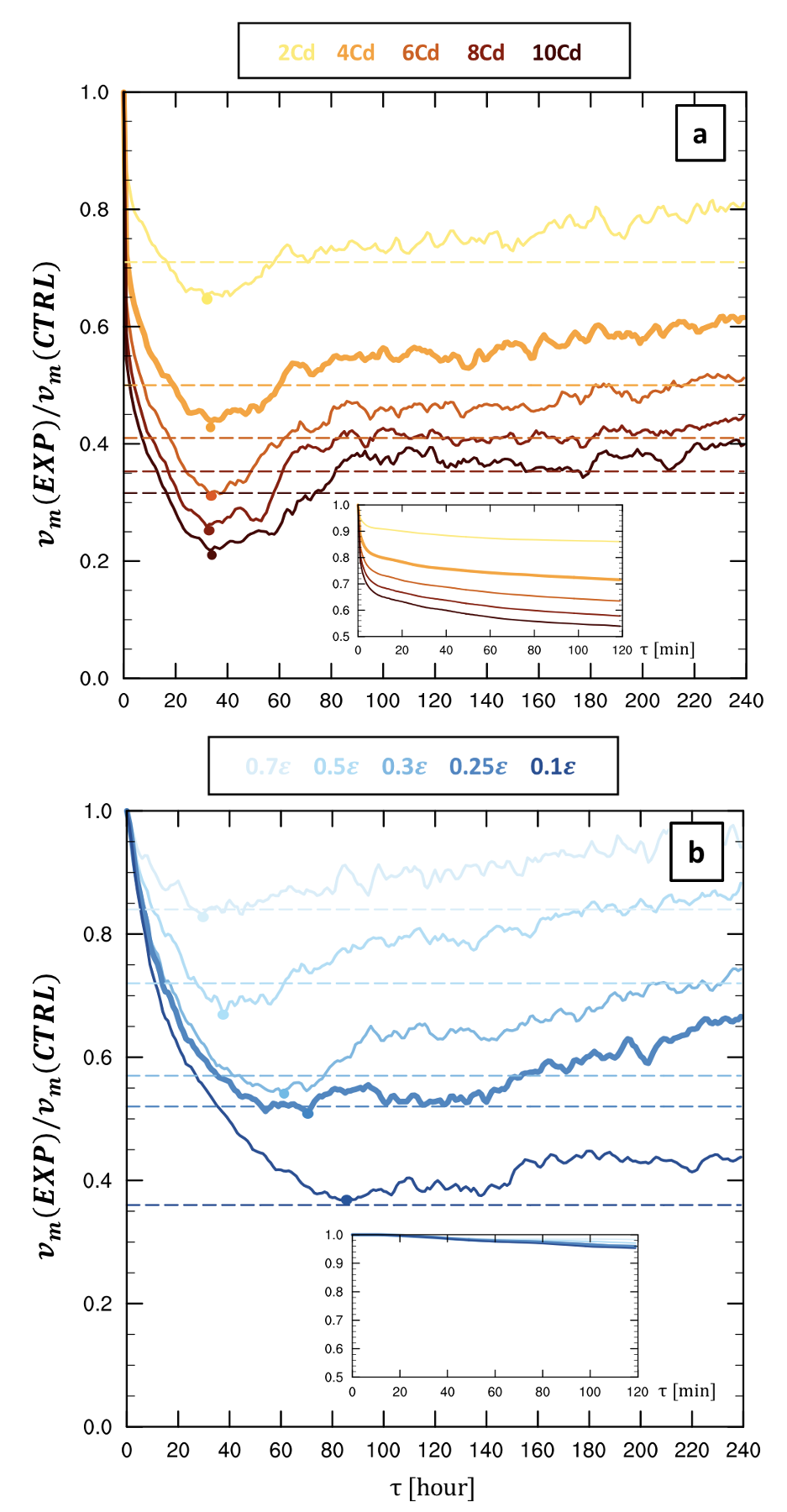}}
\caption{Temporal evolution of normalized storm intensity, $\frac{v_{{m}_{EXP}}}{v_{{m}_{CTRL}}}$, for (a) surface roughening, and (b) surface drying, with theoretical prediction (dashed line) and minimum normalized intensity (dot) for each experiment. The initial CTRL $v_m$ is 90.4 $m\,s^{-1}$ and the CTRL used in this normalization is time-dependent.  Subplots show the first 120 minutes change for each experiment. Dashed theoretical predictions are calculated from environmental $V_p$ (Eq.\eqref{predictedVm}). The $4C_d$ and $0.25\epsilon$ experiments are marked with bold lines for in-depth analysis.}\label{intensityall}
\end{figure}

\begin{figure*}[t]
\centerline{\includegraphics[width=30pc]{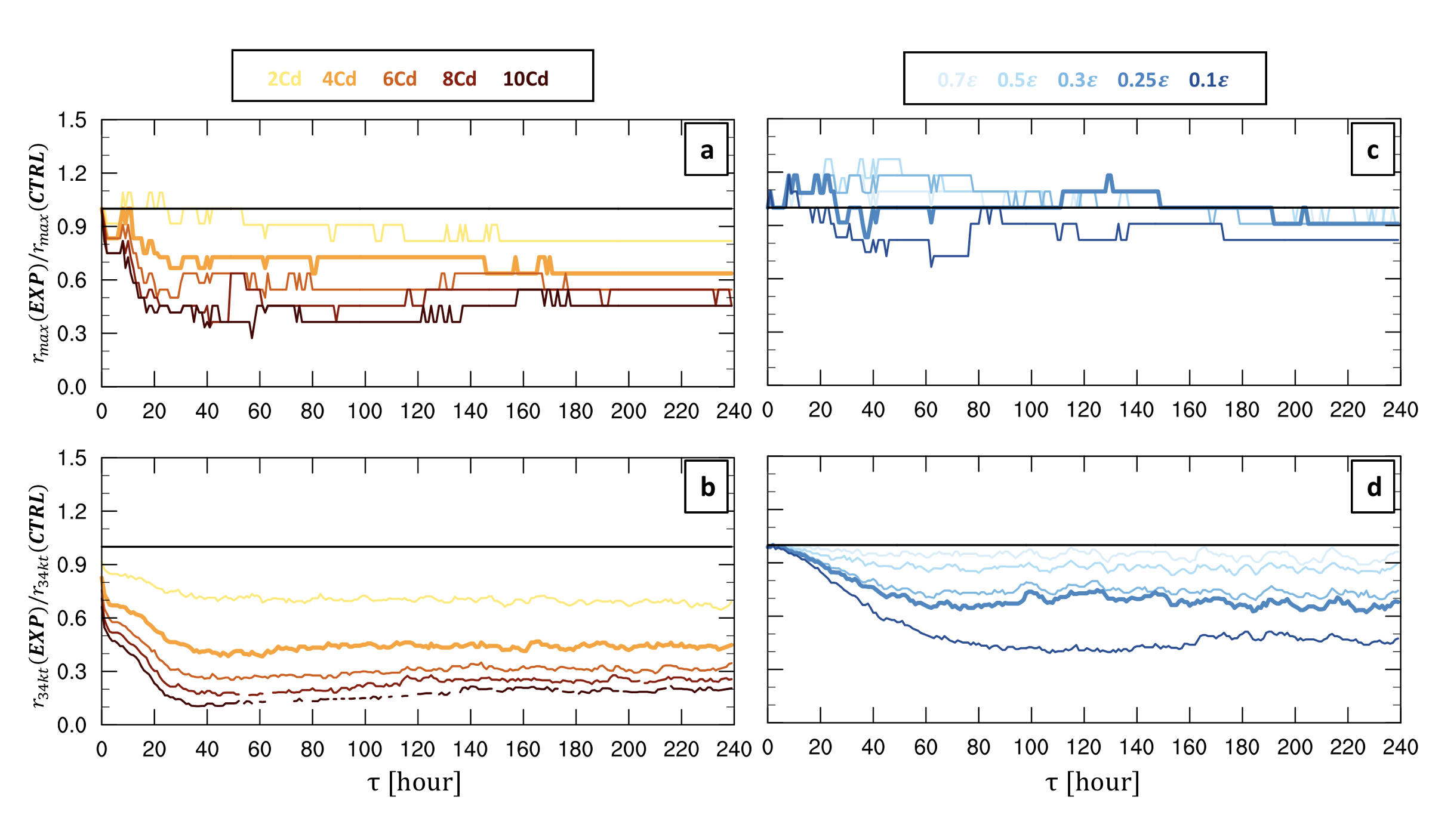}}
\caption{Temporal evolution of normalized storm size, $\frac{r_{{max}_{EXP}}}{r_{{max}_{CTRL}}}$ and $\frac{r_{{34kt}_{EXP}}}{r_{{34kt}_{CTRL}}}$, for (a-b) surface roughening, and (c-d) surface drying. The initial CTRL $r_{max}$ and $r_{34kt}$ are $33\; km$ and $258\;km$, respectively. Aesthetics as in Fig. \ref{intensityall}.}\label{sizeall}
\end{figure*}

\section{Temporal evolution of the transient responses to surface roughening and drying}\label{response1}

We first analyze the transient response of storm intensity $v_m$ relative to the CTRL (Fig.\ref{intensityall}a-b), as well as the theoretical response given by Eq.\ref{predictedVm}. Storm intensity decreases monotonically with increasing roughening or drying across all experiments. However, the transient responses to roughening vs. drying differ markedly in the intrinsic response time-scales.

For surface roughening, $v_m$ initially exhibits a very rapid decay across all experiments (subplot in Fig.\ref{intensityall}a).  Notably, there is not a transient period of intensification with roughening that has been found in previous simulations of frictional vortex spin down \citep{Montgomery2001}. $v_m$ then decays more slowly towards a minimum intensity that is smaller than the theoretical response by $\tau=40\;h$. Thereafter, $v_m$ re-intensifies to a new equilibrium that is slightly larger than the theoretical response by $\tau=80\;h$, and this equilibrium lasts at least two days. The time-scales of this transient response in $v_m$ appear to be independent of the roughening magnitude; the roughening magnitude primarily modulates the magnitude of the response.

In contrast, for surface drying, $v_m$ responds negligibly within the first hour and only weakly by $\tau=2\;h$ (subplot in Fig.\ref{intensityall}b). $v_m$ then decays slowly and quasi-exponentially towards its minimum value that is comparable to the theoretical equilibrium response. $v_m$ retains this minimum value for approximately 2 days for $0.1\epsilon$ and $0.25\epsilon$ and 1 day for $0.3\epsilon$, $0.5\epsilon$ and $0.7\epsilon$, with gradual re-intensification thereafter. The existence of this new equilibrium state follows from potential intensity theory, which predicts that the storm should tend toward a new, smaller but non-zero equilibrium, as surface enthalpy fluxes are reduced but remain non-zero \citep{Kieu2015,Chavas2017}, including both latent and sensible heat fluxes. Deeper analysis of long-term behavior at and beyond this equilibrium state is left to future work. The time-scale of the transient response monotonically increases with increased magnitude drying, again in contrast to surface roughening. For both roughening and drying, the minimum surface pressure $P_{MIN}$ increases by $\tau=40\;h$ (Supplementary Figure.1), exhibiting similar transient time scales as $v_m$. Future work might seek to analyze potential changes in the wind-pressure relationship due to surface forcing \citep[e.g.][]{Kieu2010}.

The nature of the responses in $r_{max}$ and $r_{34kt}$ (Fig.\ref{sizeall}a-b) are similar to that of $v_m$ for roughening: initial rapid decrease followed by a more gradual decrease toward a new smaller equilibrium. $r_{max}$ exhibits slightly more variation within the first 18 hours, decreasing sharply within $\tau=1\;h$ then becoming constant through $\tau=12\;h$, then briefly increasing by $\tau=18\;h$. Thereafter, $r_{max}$ decays rapidly again at a rate that increases with increased roughness. After $\tau=60\;h$, $r_{max}$ becomes relatively constant.

For drying, $r_{34kt}$ decreases slowly towards its minimum and then becomes stable (Fig.\ref{sizeall}c-d), which is similar to its $v_m$ evolution. Meanwhile, the response of $r_{max}$ is more complex and non-monotonic. Initially, $r_{max}$ remains relatively constant or slightly increases across all drying experiments up to $20\;h$. Thereafter, though, $r_{max}$ decreases with a magnitude that increases with increasing drying, eventually restoring to a radius close to the CTRL $r_{max}$. For example, for strong surface drying ($0.1\epsilon$), $r_{max}$ decreases to 70\% of its CTRL value by $\tau=40\;h$, then returns to around 85\% of the CTRL value. In contrast, $r_{max}$ remains larger than CTRL for weaker drying; for $0.25\epsilon$, $r_{max}$ gradually decreases back to the CTRL value by $40\;h$ and fluctuates around this value thereafter. 

\begin{figure}[t]
\centerline{\includegraphics[width=20pc]{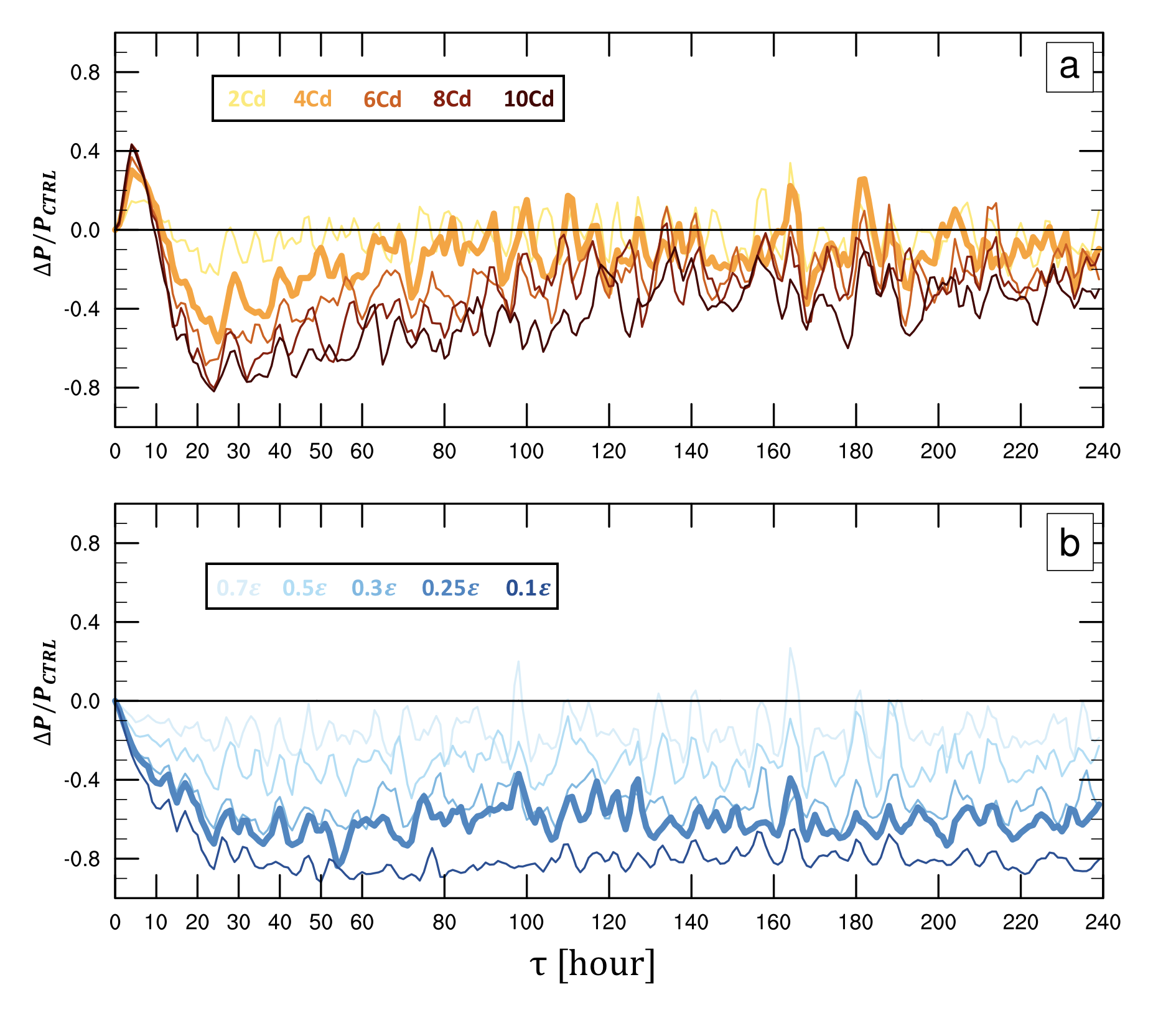}}
\caption{Temporal evolution of normalized response in total precipitation, $\frac{\triangle{P}}{P_{CTRL}}$, for (a) surface roughening and (b) surface drying. The initial CTRL $P_{CTRL}$ is 763.3$mm\,h^{-1}$. Aesthetics as in Fig. \ref{intensityall}.}\label{prcpall}
\end{figure}

Finally, we analyze the total precipitation response within $r<150 \; km$. With roughening (Fig.\ref{prcpall}a), $P$ initially increases by 15-40\%  from $\tau=0\;h$ to $\tau=5\;h$, with larger enhancement and peak value for stronger magnitude roughening. This transient precipitation enhancement consistently peaks at $\tau = 5\; h$, independent of roughening magnitude. $P$ subsequently decreases by 20-80\% of the CTRL value from $\tau=5\;h$ to $\tau=20\;h$, with larger precipitation reduction for stronger roughening. After remaining relatively constant for approximately 1 day, $P$ gradually increases back to the CTRL value and re-equilibriates after $\tau=100\;h$ across all roughening experiments. Meanwhile, for surface drying experiments, $P$ decreases by 30-90\% before reaching a new equilibrium, with larger reductions for larger magnitude drying (Fig.\ref{prcpall}b). The intrinsic time scales of this response remain constant for roughening while increase with increased drying, similar to the results above for $v_m$ and $r_{34kt}$.

These two sets of experiments demonstrate the systematic transient responses of TC intensity, size, and inner-core precipitation rate to surface roughening and drying across a range of magnitudes of each parameter. For roughening, the response magnitudes vary with roughening magnitudes, whereas the intrinsic time-scales of these responses are independent of roughening magnitudes. In contrast, for drying, the responses are generally smoother and follow a single dominant time-scale that varies with drying magnitude, with the exception of $r_{max}$ whose qualitative response exhibits a more complex dependence on the strength of drying. 

\section{Radial structure of the response in low-level wind field and precipitation field}\label{response2}
\begin{figure*}[t]
\centerline{\includegraphics[width=0.8\textwidth]{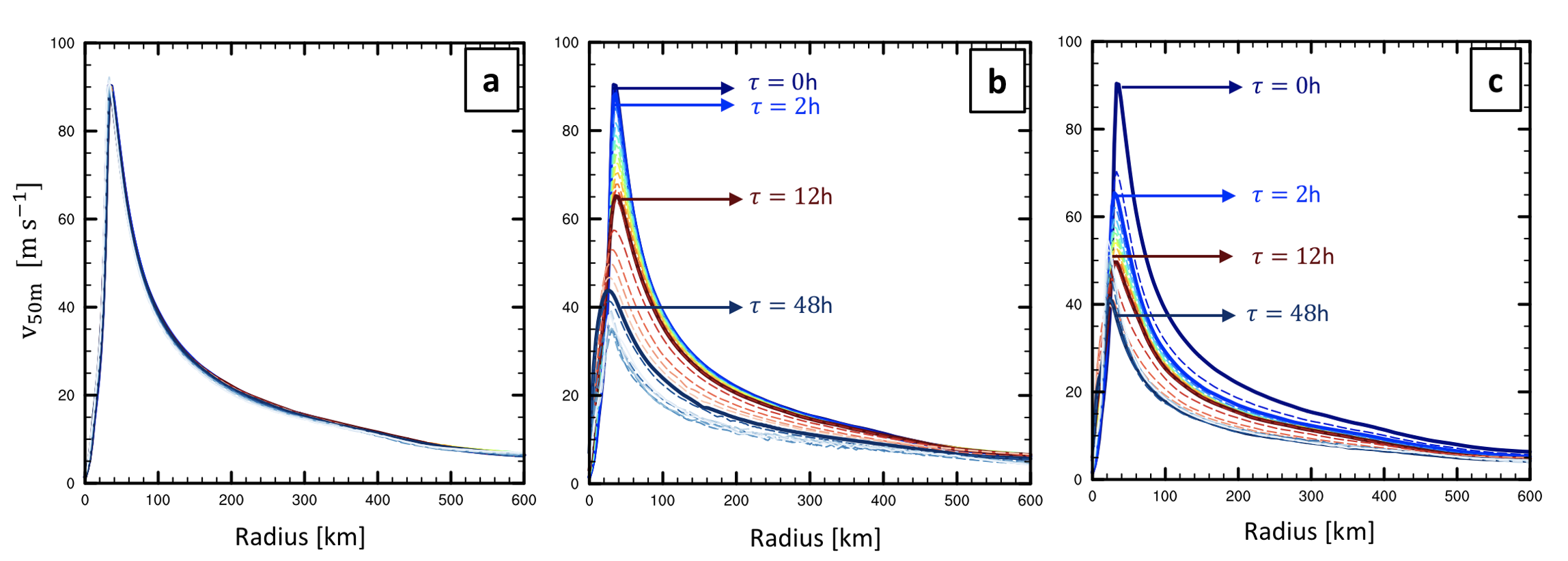}}
\caption{10-day evolution of the 50-m tangential wind profile for (a) CTRL, (b) $0.25\epsilon$ and (c) $4C_d$. Wind profiles are shown every hour for $\tau =0-12\;h$; every 6h from $\tau =12-48\;h$; and every 24h from $\tau =48$ to $240\;h$. Wind fields at $\tau=0,2,12,48\;h$ are marked and plot in solid bold lines in sequence. }\label{llwindfields}
\end{figure*}
\begin{figure*}[t]
\centerline{\includegraphics[width=0.8\textwidth]{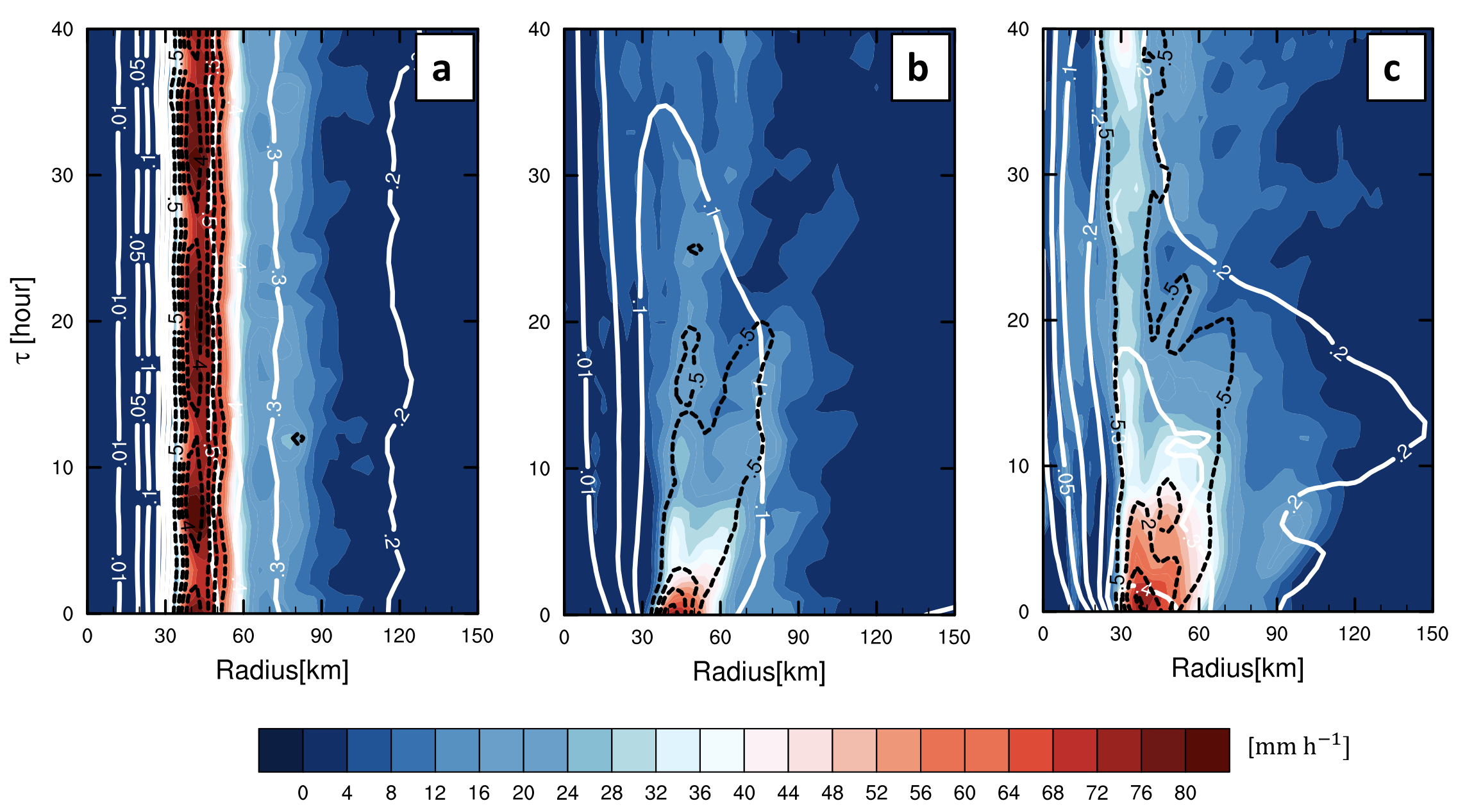}}
\caption{Hovm\"{o}ller diagram of precipitation (color), hourly-averaged $w_{2km}$ ([$m\,s^{-1}$]; dash contour) and $F_{qv}$ ([$g/kg\, m/s$]; solid contour) for a) CTRL, b) $0.25\epsilon$, and c) $4C_d$ from $\tau=0$ to $40\;h$.}\label{prcp}
\end{figure*}
We next perform an in-depth analysis of one representative experiment from each set: $0.25\epsilon$ and $4C_d$ (bold lines in Fig.\ref{intensityall}-\ref{prcpall}). We focus on the first 40 hours, during which intensity decreases rapidly and monotonically across all roughening and drying experiments. The slow longer-term re-intensification, which is likely tied to the gradual adjustment of the environment to these changes in surface properties, is generally less relevant to our primary research questions regarding landfall and so is left for future work. 

How do changes in TC intensity and size manifest themselves in the response of the radial structure of the low-level wind field? Though the nature of this weakening differs markedly between drying and roughening, the near-surface and low-level tangential wind field weakens monotonically at all radii (Fig.\ref{llwindfields}). Surface drying predominantly weakens the inner-core wind field slowly by $\tau=12\;h$, whereas the impact on the broad outer circulation decreases moving outwards to larger radii, with limited weakening beyond $r=100 \; km$ by $\tau=12\;h$ and modest weakening at $r=500\;km$ by $\tau = 48 \;h$ (Fig.\ref{llwindfields}b). In contrast, surface roughening weakens the low-level wind field at all radii strongly and rapidly within the first two hours and more slowly thereafter (Fig.\ref{llwindfields}c). Responses of the outer circulation between drying and roughening experiments also explain the evolution of $r_{34kt}$ (Fig.\ref{sizeall}d and b): for surface drying, $r_{34kt}$ changes modestly over the first 24 hours due to the negligible initial response of its outer circulation. In contrast, when the complete low-level wind field weakens rapidly due to surface roughening, wind speeds at the original $r_{34kt}$ (r=258km) also decrease well below $34\;kt$ ($17.5ms^{-1}$) rapidly. Thus, the location of $34\;kt$-wind shifts inward to much smaller radius. 

As noted in Section \ref{response1}, total inner-core precipitation responds differently to surface drying and roughening by $\tau=5\;h$: $P$ is 25\% lower for $0.25\epsilon$ and 30\% higher than CTRL for $4C_d$ (Fig.\ref{prcpall}). Here we further compare the spatiotemporal distribution of rainfall, in conjunction with surface mixing ratio fluxes $F_{qv}$ and 2-km updraft $w_{2km}$ (Fig.\ref{prcp}); the latter two variables will be discussed in the mechanistic analysis below. For surface drying, heavy precipitation exceeding $50\;mm\, h^{-1}$ quickly diminishes to below $30\;mm\, h^{-1}$ after $\tau=5\;h$ (Fig.\ref{prcp}b). By $\tau=10\;h$, precipitation primarily decreases within the inner-core, leaving a relatively constant radial extent of lighter precipitation rates less than $20\;mm\,h^{-1}$, which is consistent with the response of the low-level wind field. 
\begin{figure*}[t]
\centerline{\includegraphics[width=0.8\textwidth]{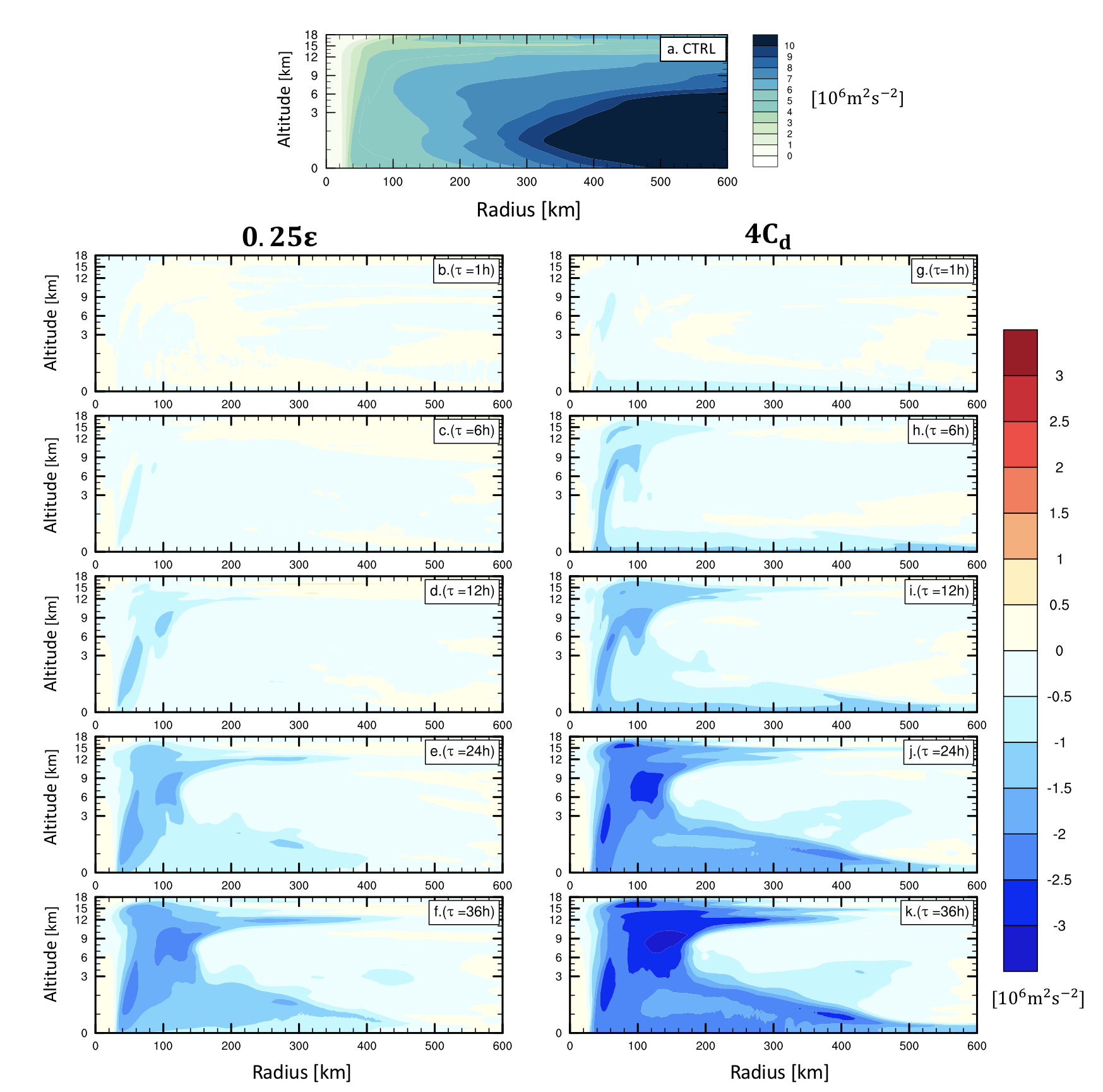}}
\caption{CTRL angular momentum field (a) and response of angular momentum ($\triangle{M}$) at $\tau=1, 6, 12, 24$ and $36\;h$ for the $0.25\epsilon$ (b-f) and $4C_d$ (g-k). Values calculated as 1-hour averages centered on given time. CTRL shows 1-hour average centered at $\tau = 1 \; h$. }\label{Mevolution}
\end{figure*}
For surface roughening, heavy precipitation expands both inwards and outwards over a wider area from $r=30$ to $60\;km$ during $\tau < 5 \;h$, whereas in CTRL, it is confined to within $r=40-55\;km$ (Fig.\ref{prcp}a and c). Meanwhile, the region of lighter precipitation rates expands to larger radii beyond $r=90\;km$. After $\tau=5\;h$, the region of heavier precipitation disappears, yet lighter precipitation persists out beyond $r=90\;km$. After $\tau=10\;h$, a concentrated annulus of higher precipitation gradually reemerges associated with the redevelopment of the eyewall at $r=30\;km$. Deeper analysis of precipitation changes is provided in Section \ref{mechanism}\ref{precipitation}.

\begin{figure*}[t]
\centerline{\includegraphics[width=0.8\textwidth]{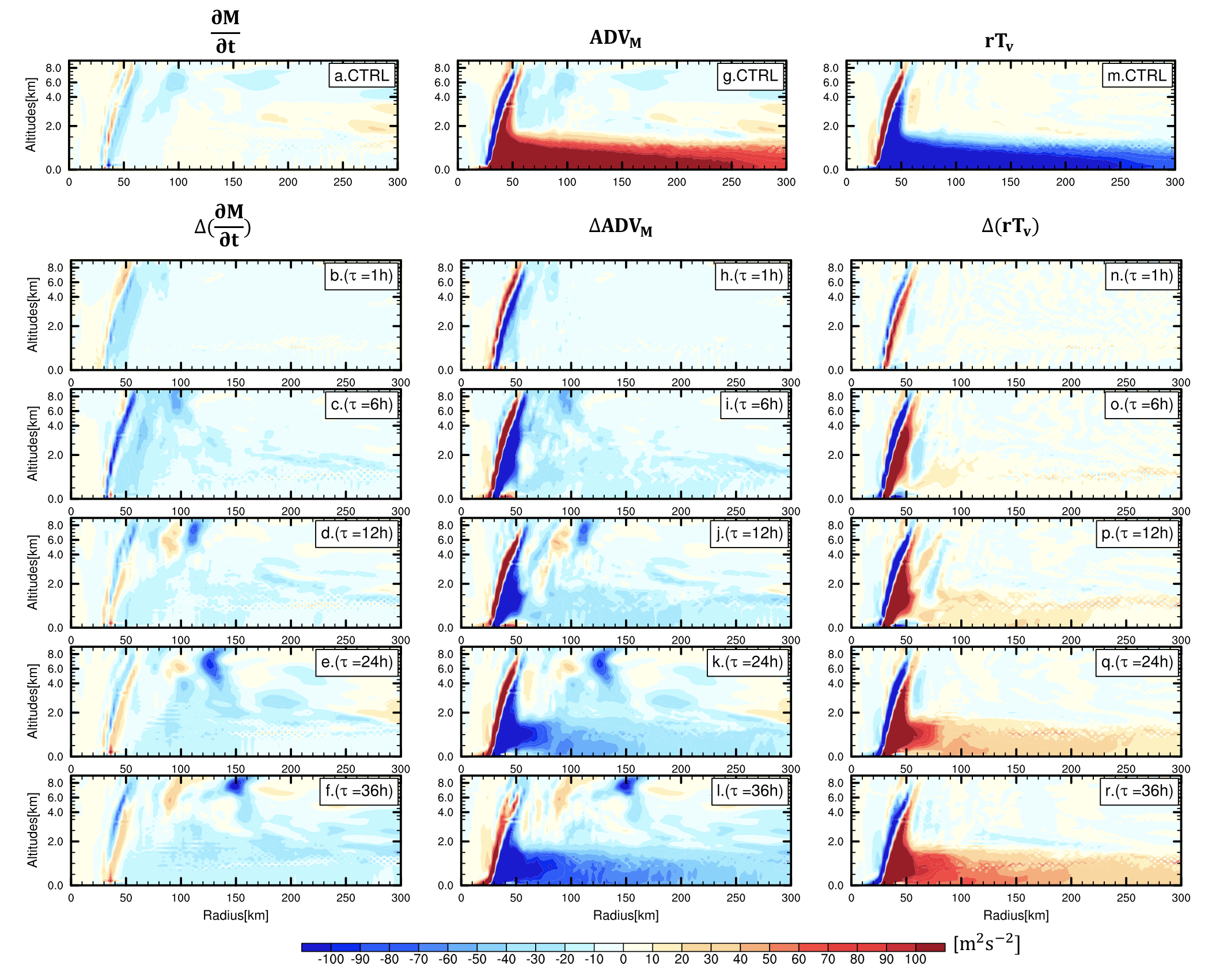}}
\caption{CTRL angular momentum budget and evolution of the angular momentum response budget for $0.25\epsilon$. Terms include tendency (a-f), advection (g-l), and turbulent mixing (m-r). Values calculated as 1-hour averages centered on given time. CTRL shows 1-hour average centered at $\tau = 1 \; h$.}\label{MbgtFLX}
\end{figure*}

\section{Physical mechanisms}\label{mechanism}

We next seek a detailed mechanistic understanding of the responses to roughening and drying characterized above. We again focus on our two representative experiments: $0.25\epsilon$ and $4C_d$. Since the responses to surface drying follow a single dominant time-scale as described in Section \ref{response1} and \ref{response2}, we begin our analysis with the $0.25\epsilon$ experiment.

\subsection{Low-level wind field}

\begin{figure*}[t]
\centerline{\includegraphics[width=0.8\textwidth]{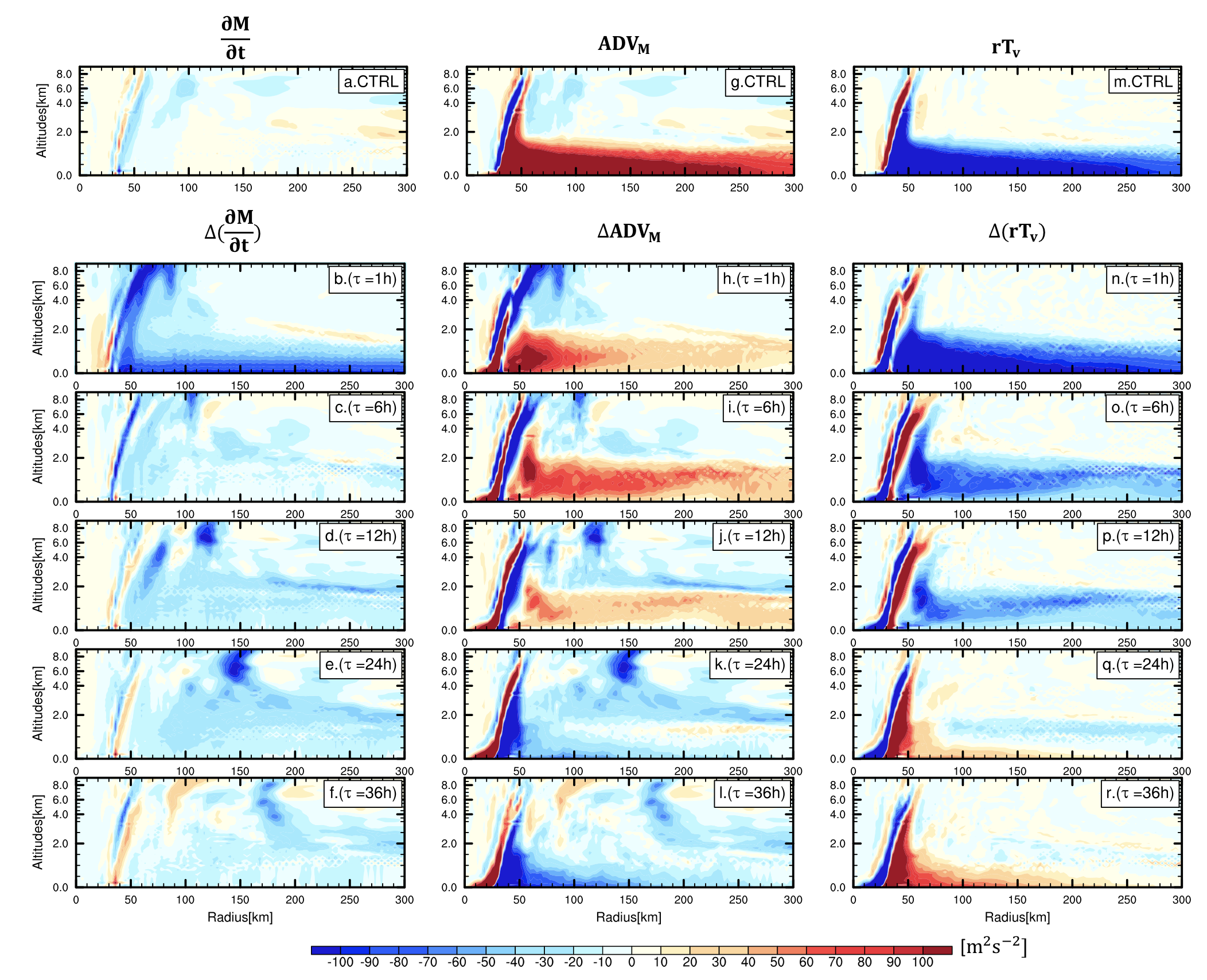}}
\caption{Same as Figure \ref{MbgtFLX} but for $4C_d$.}\label{MbgtDRG}
\end{figure*}

To understand the responses of low-level wind field, we quantify the response of absolute angular momentum, $\Delta M$ (Fig.\ref{Mevolution}), and analyze the physical processes controlling $\Delta M$ in for $0.25\epsilon$ (Fig.\ref{MbgtFLX}) and $4C_d$ (Fig.\ref{MbgtDRG}). We plot pointwise differences from CTRL rather than side-by-side comparisons to highlight the detailed structure of the responses.

\subsubsection{Response to surface drying ($0.25\epsilon$)}
For surface drying, $M$ changes minimally throughout the whole domain during the first hour (Fig.\ref{Mevolution}b). By $\tau = 6 \; h$, $M$ begins to decrease within the eyewall aloft, while the near-surface $M$ remains comparable to the CTRL (Fig.\ref{Mevolution}c). By $\tau = 12 \; h$, this response extends upwards into the outflow, whereas within the boundary layer it remains confined to near the radius of maximum wind(Fig.\ref{Mevolution}d). During $\tau=12-24 \; h$, the response extends radially outward at low levels (Fig.\ref{Mevolution}e). By $\tau=36 \; h$, the response pattern is similar to $24 \; h$ but with stronger magnitude: $(\triangle{M})_{max}$ reaches $-1.5\times  \; 10^6 m^2\,s^{-1}$ near the outflow region (Fig.\ref{Mevolution}f). As a result, the low-level wind field weakens slowly and primarily within the inner region during the first 2 hours, as found in Section \ref{response2}.

We use Eq.\ref{MbdgtDiff} to quantify how changes in advection and turbulent mixing, including surface frictional dissipation, control the response tendency, $\Delta(\frac{\partial M}{\partial t})$, and thus $\Delta M$ (Fig.\ref{MbgtFLX}). We first $\triangle({-u\frac{\partial M}{\partial r}})$ and $\triangle({-w\frac{\partial M}{\partial z}})$ together into a total advection response term, $\triangle{ADV_M}$. Within the inner-region boundary layer, the dominant source term of $M$ is due to advection (Fig.\ref{MbgtFLX}g) and the dominant sink term is due to turbulent dissipation by surface friction (Fig.\ref{MbgtFLX}m); the latter depends on both surface drag coefficient and near-surface total wind speed (Eqs.\ref{Eq_Cd_z0}-\ref{Eq_Cd_turb}). The evolution of $\Delta(\frac{\partial M}{\partial t})$ is controlled primarily by the decreasing advective $M$ source near the surface (Fig.\ref{MbgtFLX}h-l). Changes in $M$ loss by turbulent dissipation, $\triangle{(rT_v)}$, depends solely on changes in near-surface wind speed since $C_d$ is fixed. Thus, for $\tau=0-2\;h$, during which the wind field remains nearly constant (Fig.\ref{llwindfields}b), near-surface $\triangle{(rT_v)}$ changes little relative to CTRL (Fig.\ref{MbgtFLX}n). During this time, $\triangle (\frac{\partial M}{\partial t})$ remains near zero initially except for the slight radial expansion near eyewall (Fig.\ref{MbgtFLX}b). Through $\tau=36\;h$, the turbulent sink term monotonically increases towards smaller negative values (i.e., a weaker sink of $M$) due to the slowly-weakening primary circulation (Fig.\ref{MbgtFLX}n-r). Meanwhile, the advective $M$ source is gradually decreasing as well (Fig.\ref{MbgtFLX}h-l). However, the advective source is consistently slightly smaller in magnitude than the frictional sink, yielding a monotonically weakening storm (Fig.\ref{MbgtFLX}b-f). Beyond $\tau=36\;h$, the weakened frictional $M$ sink and advective $M$ source begin to more closely balance each other as the storm approaches a new, weaker equilibrium. 

For deeper analysis of the boundary layer response, we integrate each of the two dominant source/sink terms within the cylinder bounded by $r\le 300 \; km$ and $z \le 2 \; km$, i.e.
\begin{equation}\label{Equation_integrateMbgt}
\int_{0}^{z=2km}\int_{0}^{r=300km}{X} rdrdz
\end{equation}
where $X$ represents the response tendency ($\triangle(\frac{\partial M}{\partial t})$), the frictional sink response ($\triangle{(rT_v)}$), the radial and vertical advection responses (${\triangle{(-u\frac{\partial M}{\partial r})}}$ and  ${\triangle{(-w\frac{\partial M}{\partial z})}}$, respectively). The results for both drying and roughening experiments are shown in Fig.\ref{trend}. The advection response is generally dominated by the radial component as expected, as this component represents the dominant source of advection for the system within the boundary layer. Comparison among source and sink terms over time (Eq.\ref{Equation_integrateMbgt}; Fig.\ref{trend}a) confirms that the inward advective $M$ source is consistently insufficient to balance the frictional dissipation sink, resulting in negative values of $\frac{\partial M}{\partial t}$, i.e. the gradual loss of $M$ and thus the weakening of low-level wind field in the inner core with time. We note that this two-region vortex behavior, in which the inner-core intensity changes independent of a relatively stable outer circulation, aligns closely with the physical wind structure model of \citet{Chavas+Lin2016}.

\subsubsection{Response to surface roughening ($4C_d$)}
For surface roughening, $M$ initially decreases rapidly near the surface at all radii (Fig.\ref{Mevolution}g), and this signal subsequently extends upward within the eyewall during the first hour. By $\tau=12 \; h$, the reduction of $M$ extends radially outward within the boundary layer and in the outflow above the eyewall, with $(\triangle{M})_{max}$ exceeding $-2\times 10^6 m^2\,s^{-1}$ (Fig.\ref{Mevolution}h-i). During $\tau=12-36 \; h$, $M$ decays through the depth of the vortex, with largest magnitude changes in the outflow region where $(\triangle{M})_{max}$ exceeds $-3\times 10^6 \;  m^2\,s^{-1}$ (Fig.\ref{Mevolution}j-k). This reduction of $M$ near the surface at all radii translates directly to the strong and rapid weakening of the complete low-level tangential wind field (Fig.\ref{llwindfields}c).
\begin{figure}[t]
\centerline{\includegraphics[width=20pc]{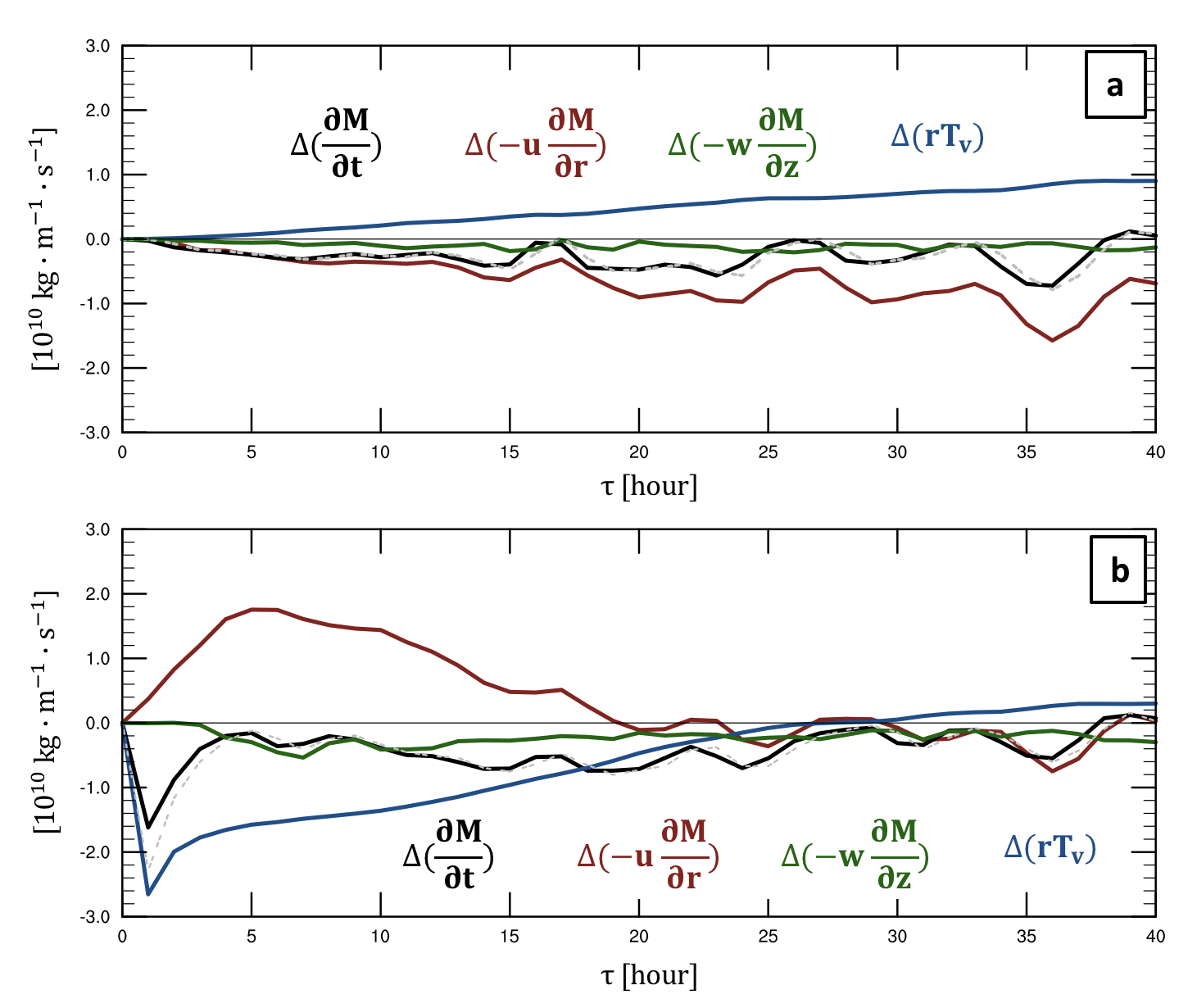}}
\caption{Temporal evolution of the integrated angular momentum budget responses (Eq. \eqref{Equation_integrateMbgt}) for (a) $0.25\epsilon$ and (b) $4C_d$. Total integrated tendency (black) and sum of the three budget terms (dash grey) are shown.}\label{trend}
\end{figure}

\begin{figure}[t]
\centerline{\includegraphics[width=20pc]{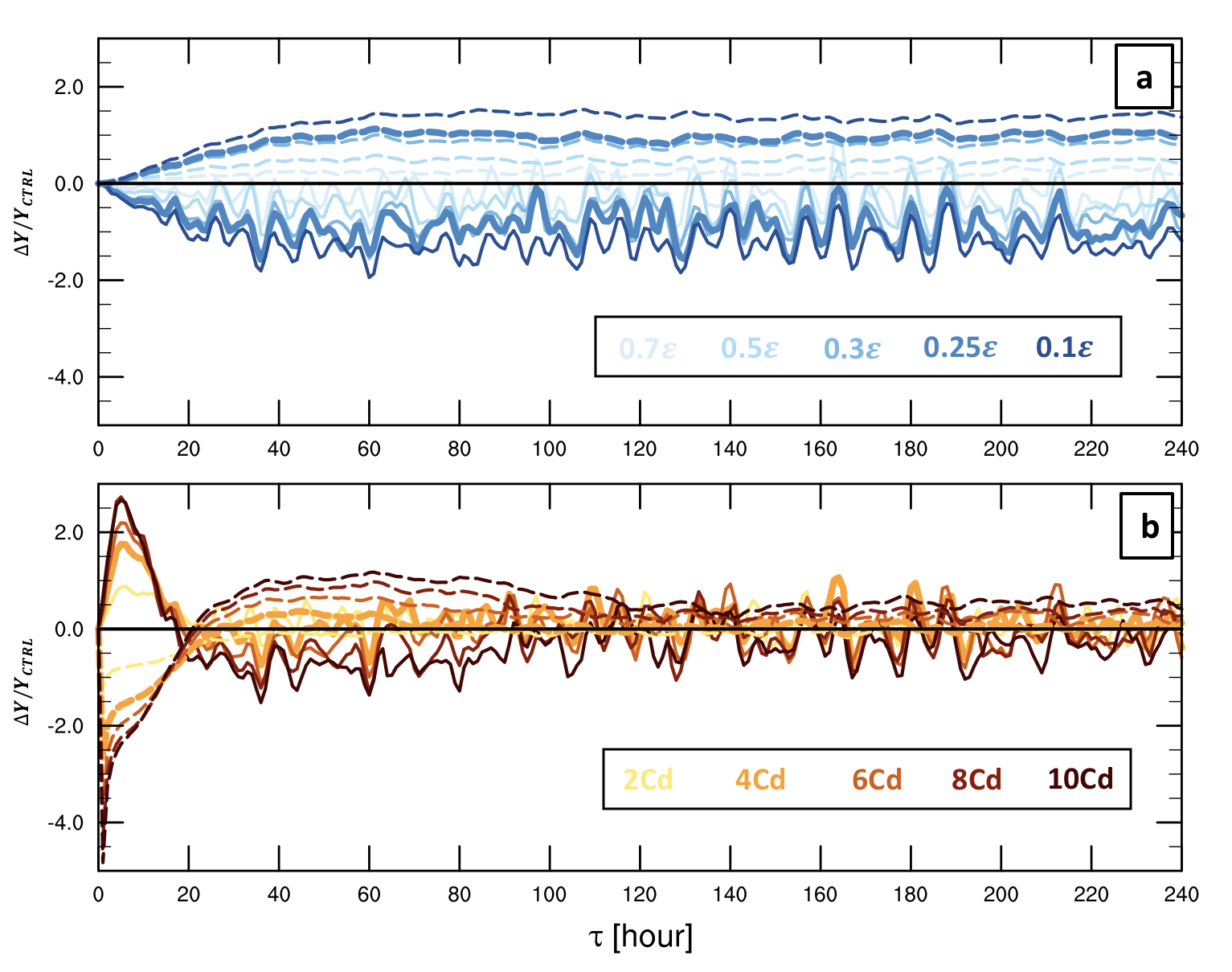}}
\caption{Temporal evolution of integrated responses of radial advective $M$ source (solid lines; $\triangle{(-u\frac{\partial M}{\partial r})}$ in Fig.\ref{trend}) and turbulent $M$ sink (dash lines; $\triangle{(rT_v)}$ in Fig.\ref{trend}), calculated by Eq.\eqref{Equation_integrateMbgt} and divided by its CTRL value as $\frac{\Delta Y}{Y_{CTRL}}$ for (a) surface drying experiments and (b) surface roughening experiments, $Y$ represents budget term $-u\frac{\partial M}{\partial r}$ or $rT_v$. Presentation as in Fig. \ref{prcpall}. } \label{trendAll1}
\end{figure}
We again analyze the evolution of $\Delta M$ via budget analysis (Fig.\ref{MbgtDRG}). Initially ($\tau < 1\;h$), boundary-layer $\frac{\partial M}{\partial t}$ decreases rapidly at all radii (Fig.\ref{MbgtDRG}b) due to the greatly enhanced $M$-sink (Fig.\ref{MbgtDRG}n), which is forced by the increased drag coefficient $C_d$. This enhanced sink exceeds the moderately-enhanced advective $M$ source within the inner core $r<100 \; km$ (Fig.\ref{MbgtDRG}h). By $\tau=12 \; h$, the storm weakens more slowly, with $\Delta (\frac{\partial M}{\partial t})$ decreasing in magnitude to $-20\; m^2\,s^{-2}$ (Fig.\ref{MbgtDRG}c-d). This decrease results from the reduction in $\triangle{(rT_v)}$ (Fig.\ref{MbgtDRG}o-p) due to the strongly-weakened low-level wind speeds (Fig.\ref{llwindfields}c). Meanwhile, the initial enhancement of $ADV_M$ also decays gradually during this period (Fig.\ref{MbgtDRG}i-j), thus coming closer to balancing the frictional $M$-sink. Finally, after $\tau=24\; h$, $\frac{\partial M}{\partial t}$ gradually decreases towards zero in much of the inner domain (Fig.\ref{MbgtDRG}e-f), indicating that the vortex is approaching a new equilibrium. At low levels, the advective source response changes sign and becomes slightly negative (i.e. a weaker source; Fig.\ref{MbgtDRG}k-l), while the turbulent sink response becomes slightly positive (i.e. a weaker sink; Fig.\ref{MbgtDRG}q-r). This final behavior reflects the much weaker primary and secondary circulations overlying a rougher surface. 

Finally, we analyze the temporal evolution of the integrated response of $M$ sources and sinks to identify the dominant physical process for the boundary layer $M$ tendency (Fig.\ref{trend}b). By $\tau=2\;h$, the frictional sink is enhanced significantly more than the advective source, and this imbalance causes $\Delta (\frac{\partial M}{\partial t})$ to decrease rapidly. By $\tau=5\;h$, the advective source response reaches its maximum magnitude, though it is never sufficient to fully balance the surface frictional dissipation source. Thus, $\triangle (\frac{\partial M}{\partial t})\leq 0$ before reaching a new equilibrium. 

It is insightful to compare the evolutions of the integrated responses of the advective $M$ source and frictional $M$ sink terms shown in Fig.\ref{trend} across all drying and roughening experiments (Fig.\ref{trendAll1}). Each integrated response is divided by its CTRL value, $\frac{\Delta Y}{Y_{CTRL}}$, where $Y$ represents the radial advective $M$ source or frictional $M$ sink. All surface drying experiments produce qualitatively similar responses that are amplified with stronger drying, but with time scales that increase with drying magnitudes. All surface roughening experiments also have similar qualitative response evolutions that are amplified with stronger roughening, but with constant time scales for the peak enhancement in advective $M$ source by $\tau = 5\;h$ and the re-equilibration of both source and sink terms after $\tau = 24\;h$. The distinct time-scales of the storm responses characterized in Section \ref{response1} appear intimately tied to the responses of these underlying source and sink terms in each experiment. 

\begin{figure*}[t]
\centerline{\includegraphics[width=0.9\textwidth]{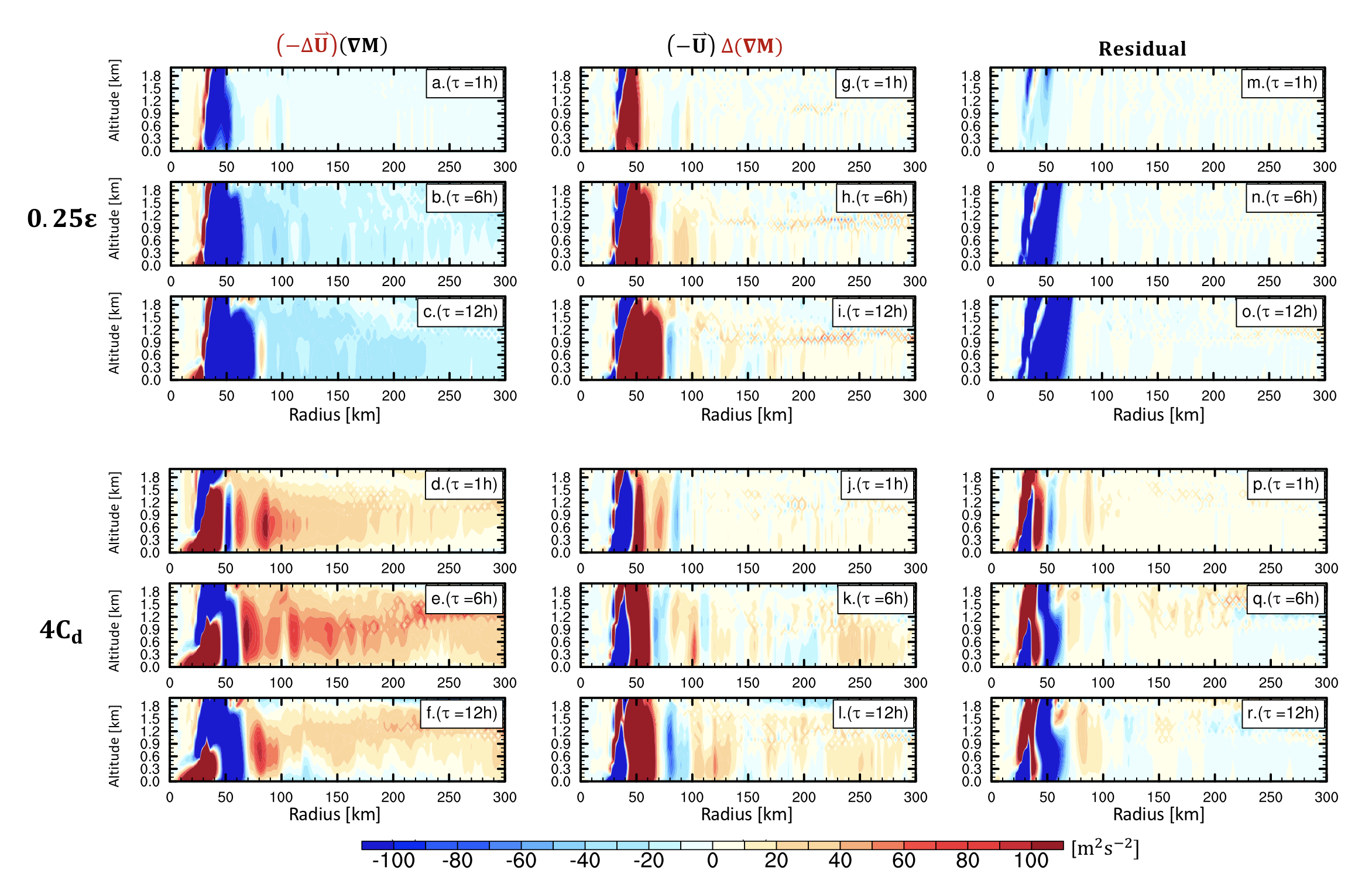}}
\caption{Linear decomposition of $\triangle{ADV(M)}$ (Fig.\ref{MbgtFLX} and Fig.\ref{MbgtDRG} middle columns) for the first 12 hours of $0.25\epsilon$ and $4C_d$. Plot displays overturning response (a-f), vortex component (g-l) and nonlinear residual (m-r).}\label{advcomp}
\end{figure*}

\begin{figure*}[t]
\centerline{\includegraphics[width=0.8\textwidth]{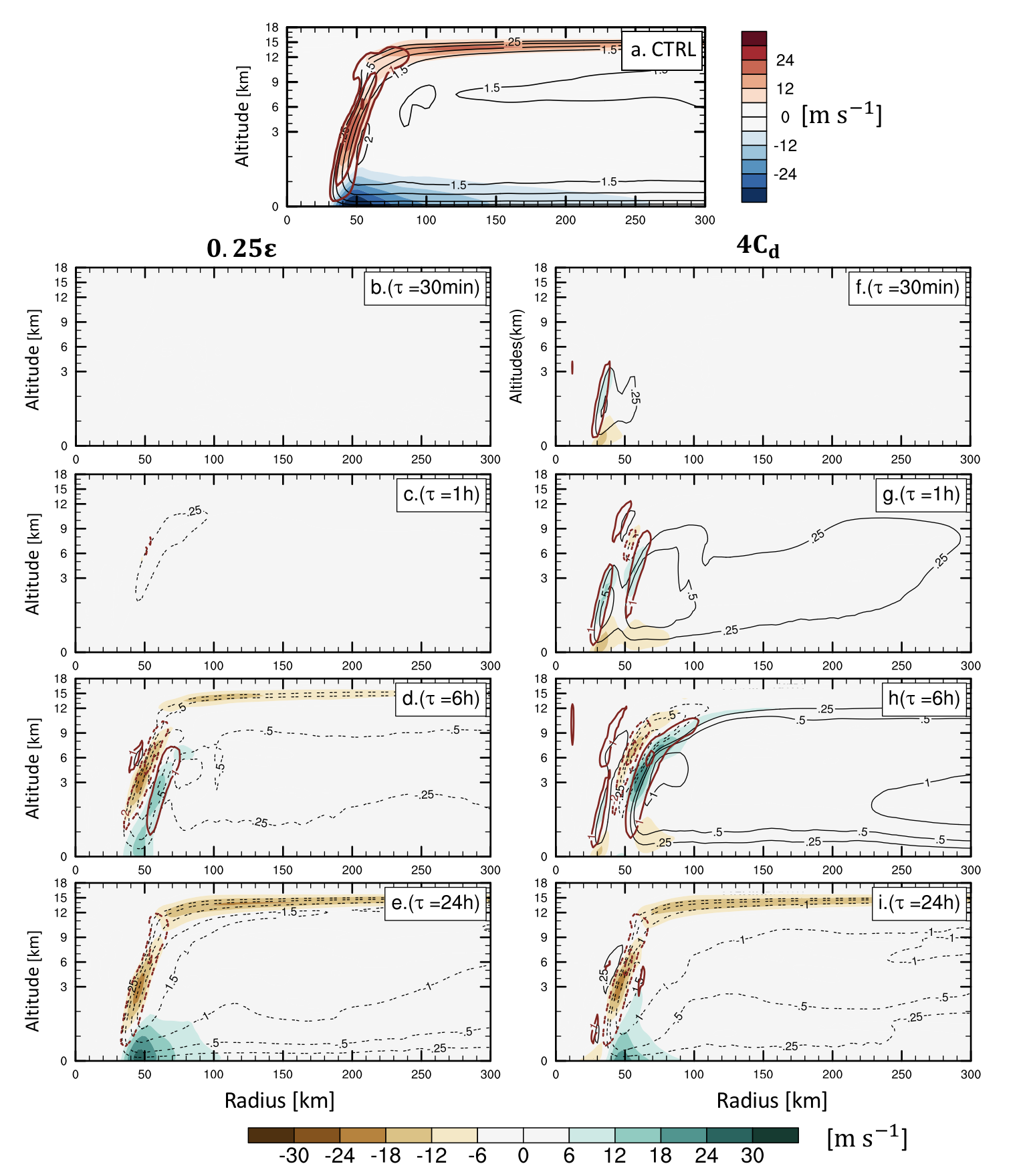}}
\caption{(a). CTRL mass stream function $\psi$ ([$10^9 kg\,s^{-1}$]; black contour), radial wind $u$ (color) and vertical velocity ${w}$ ([$m\,s^{-1}$]; red contour) at $\tau = 1 \;h$. Temporal evolution of the responses of mass stream function $\triangle{\psi}$, radial wind $\triangle{u}$ and vertical velocity $\triangle{w}$ to $0.25\epsilon$ (b-e) and $4C_d$ (f-i). $\tau=30\;min$ shows true snapshot rather than the hourly average.}\label{cir}
\end{figure*}

\begin{figure*}[t]
\centerline{\includegraphics[width=0.8\textwidth]{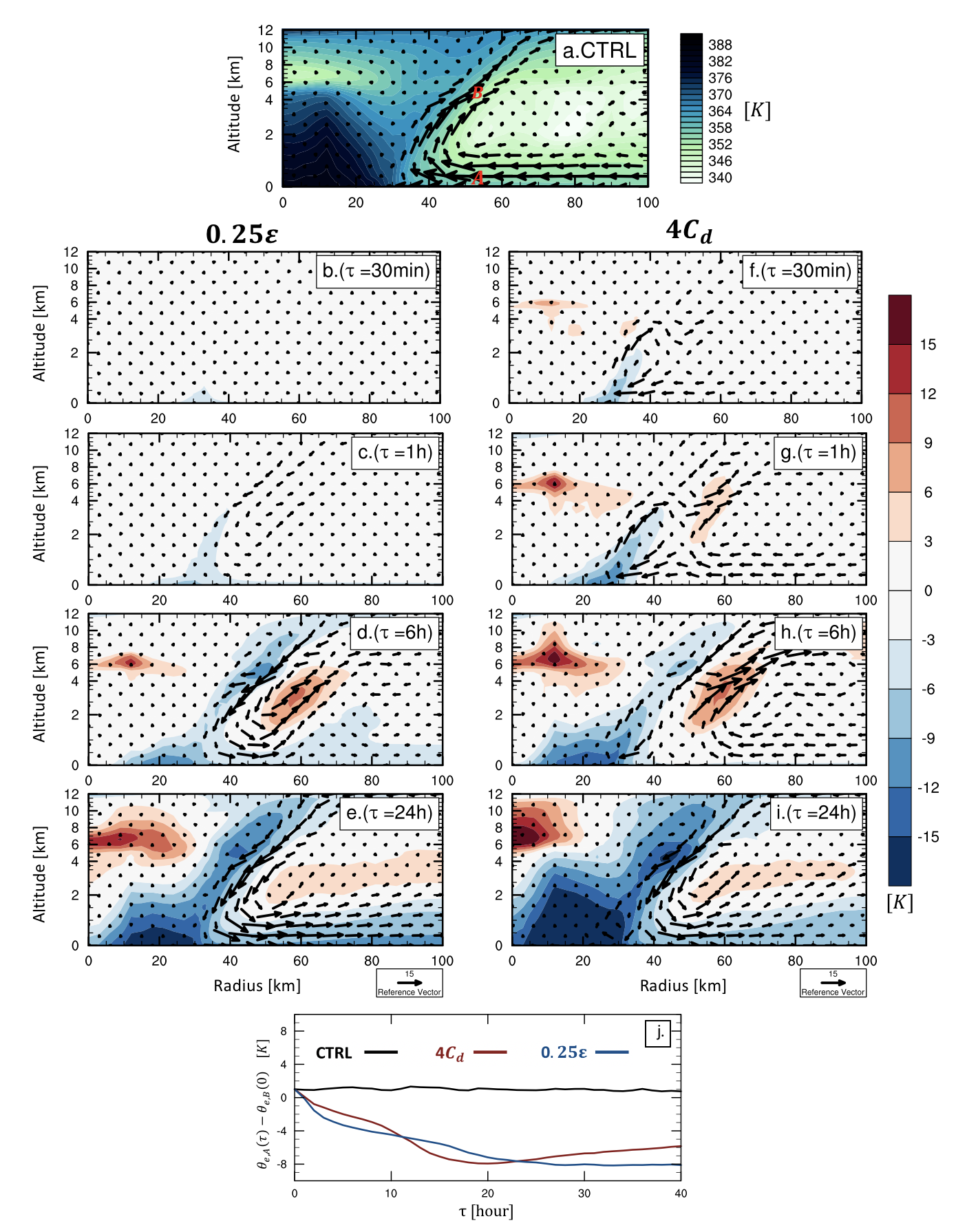}}
\caption{(a). Equivalent potential temperature, $\theta_e$, and r-z flow field, $\vec{U}$ for CTRL at $\tau = 1 \; h$. Temporal evolution of the response of equivalent potential temperature, $\Delta \theta_e$, response of r-z flow field, $\Delta \vec{U}$, for (b-e) $0.25\epsilon$ and (f-i) $4C_d$. In (a), marker B denotes location of maximum vertical velocity within the eyewall at $\tau=0\;h$ ($(r,z)=(54\;km, 5.24\;km)$), and marker A denotes near-surface location beneath B ($(r,z)=(54\;km, 0.05\;km)$). (j) Temporal evolution of measure of eyewall stability, defined as $\theta_{e,A}(\tau)-\theta_{e,B}(0)$, for each experiment. Values calculated as 1-hour averages centered on given time except for $\tau=30\;min$, which shows true snapshot.}\label{thetaE}
\end{figure*}
\subsubsection{Contrasting boundary-layer responses to roughening and drying}

Ultimately, the responses of the underlying processes controlling the primary circulation (i.e. vortex) described above (Fig.\ref{MbgtFLX} and \ref{MbgtDRG} middle columns) are intimately linked to the responses of the secondary overturning circulation.

To better understand their interdependent dynamics, we first linearly decompose the advective term to quantify the relative contributions of changes in the primary circulation and secondary circulation during the first 12 hours. This decomposition is given by:
\begin{equation}\label{ADVanalysis}
\triangle({\vec{U} \cdot \nabla M})=\triangle{\vec{U}}\cdot \nabla M+ \vec{U}\cdot\Delta(\nabla M)+ \Delta \vec{U}\cdot\Delta(\nabla M)
\end{equation} where $\vec{U}=(u,w)$ and $\nabla M =(\frac{\partial M}{\partial r}, \frac{\partial M}{\partial z})$. The first term on the RHS of Eq.\ref{ADVanalysis} is the overturning response, associated with the linear response due to the change in $\vec{U}$ holding the CTRL vortex ($\nabla M$) constant. The second term is the vortex response, associated with the linear response of $\nabla M$ holding the CTRL overturning ($\vec{U}$) constant. The final term is the non-linear residual, which quickly becomes large as the responses of the primary and secondary circulations both become strong. We focus on the contribution of each term within the boundary layer in the inner core of the storm within the initial 12 hour window (Figures \ref{MbgtFLX} and \ref{MbgtDRG}).

For both drying and roughening, $\triangle{ADV_M}$ within the boundary layer is primarily controlled by the response of the overturning circulation. For surface drying, the weakened overturning circulation (Fig.\ref{advcomp}a-c) leads to the reduction in $ADV_M$ (Fig.\ref{MbgtFLX}h-l). For roughening, the enhancement of $ADV_M$ (Fig.\ref{MbgtDRG}h-l) is dominated by the stronger overturning circulation (Fig.\ref{advcomp}d-f). The vortex responses to either surface drying or roughening are concentrated near the eyewall (Fig.\ref{advcomp} g-l), where $\frac{\partial M}{\partial r}$ is largest. The non-linear residual term increases near the eyewall with time (Fig.\ref{advcomp}m-r). We do not address mechanisms associated with changes in the vortex radial structure here, but this may be an important an avenue of future work, particularly for understanding changes in $r_{max}$; one simple analysis in this direction is provided in the next subsection.  

What controls these contrasting responses of the overturning circulation to each type of surface forcing? The overturning flow response and concurrent thermodynamic response for each experiment are shown in Figures \ref{cir} and \ref{thetaE}, respectively.

For surface drying, $ADV_M$ weakens monotonically as the overturning circulation gradually decays (Fig.\ref{cir}b-e), which starts from the eyewall aloft and reflects the stabilization of the eyewall. This eyewall stabilization process may be quantified via the thermodynamic response of equivalent potential temperature, $\Delta \theta_e$ (Fig.\ref{thetaE}). Specifically, we define a simple measure of eyewall stability as $\theta_{e,A}(\tau)-\theta_{e,B}(0)$, where B denotes the location of maximum vertical velocity in the eyewall at $\tau=0\;h$, A is the near-surface location beneath B (A and B marked in Fig.\ref{thetaE}a). In CTRL, $\theta_e$ is nearly conserved moving from A to B, as would be expected within the strongly convecting eyewall (Fig.\ref{thetaE}j). In the drying experiment where low-level wind speeds remain relatively constant by $\tau=6\;h$, $\theta_e$ decreases in the boundary layer due to the direct reduction of surface moisture availability (Eq.\ref{Eq_qvflux}, Fig.\ref{thetaE}b-d). The advective response $\Delta(-u\frac{\partial \theta_e}{\partial r})$ does not contribute to this reduction in $\theta_e$, as the advective import of low $\theta_e$ air is reduced (which would act to increase $\theta_e$ if acting alone) by the gradually weakened overturning circulation (Supplementary Figure 2). Because $\theta_e$ is reduced specifically within the boundary layer and near the surface beneath the original eyewall, the convective instability of boundary layer parcels near the eyewall is reduced. This boundary-layer $\theta_e$ reduction gradually stabilizes the eyewall (Fig.\ref{thetaE}j), and thus causes the overturning circulation to gradually weaken aloft within the eyewall first (Fig.\ref{thetaE}c) and then to extend downwards into the boundary layer (Fig.\ref{thetaE}d). At $\tau=6\;h$, a slightly stronger updraft forms outside the original eyewall at $r=50\;km$ while the weakening of the overturning circulation extends through the outflow out to larger radii, with $(\triangle{\psi})_{max}\approx-1.5\times 10^9 kg\,s^{-1}$ (Fig.\ref{cir}d). The formation of this new updraft suggests an outward shift of the original eyewall, as boundary-layer air, now with a reduced $\theta_e$, can only ascend at a larger radius where mid-level $\theta_e$ values are initially lower (Fig.\ref{thetaE}a). This is perhaps a simple thermodynamic explanation for why the eyewall and the radius of maximum wind typically expand outwards with weakening. By $\tau=24 \;h$, the temporarily enhanced updraft column gradually weakens, and the overturning circulation continues to weaken throughout the entire vortex (Fig.\ref{cir}e and Fig.\ref{thetaE}e). As the overturning circulation continues to weaken, the advective response $\Delta(-u\frac{\partial \theta_e}{\partial r})$ is increasingly positive (i.e. the import of low-$\theta_e$ air is reduced), which again indicates that the reduction in $\theta_e$ is driven by the reduction in surface fluxes rather than the advective response.

For surface roughening, $ADV_M$ is initially ($\tau \le 1\;h$) enhanced by the strengthened near-surface inflow due to Ekman balance adjustment. An enhanced frictional drag weakens the tangential wind speed and thus weakens the outward forces (Coriolis and centrifugal forces), resulting in an inward force imbalance \citep{Anthes1974} and thus a stronger near-surface inflow (Fig.\ref{cir}f-g). This stronger inflow first enhances the inner-core overturning circulation, though this frictional enhancement is confined within a relatively shallow layer below $z=4\;km$ (Fig.\ref{cir}f). By $\tau=6\;h$, the overturning circulation is enhanced throughout much of the vortex out to $r=300\;km$, with two maxima in $\triangle{\psi}$ and two enhanced updraft columns on either side of the original eyewall (Fig.\ref{cir}g-h). Though enhanced inflow temporarily enhances inward advection of $M$ during the first 12 hours, stronger frictional dissipation rapidly consumes $M$ near the surface. Thus, this strengthened overturning circulation transports reduced-$M$ fluid near the surface inwards and out of the boundary layer, thereby gradually spinning down the vortex aloft through the depth of the troposphere. This overturning circulation and spin-down process is analogous to the "stirred tea cup" model in \cite{Holton2004}.

While Ekman balance adjustment rapidly enhances the overturning circulation dynamically, the reduction in surface moisture fluxes associated with the weakened near-surface circulation acts to gradually stabilize the eyewall thermodynamically. The latter eventually dominates, causing the overturning circulation to weaken and hence reduces the inward advection of angular momentum within the boundary layer (Fig.\ref{cir}i). Relative to CTRL (Fig.\ref{thetaE}a), $\theta_e$ decreases first within the boundary layer near the eyewall, where the frictionally-reduced surface wind speed reduces $F_{qv}$ (Eq.\ref{Eq_qvflux}, Fig.\ref{thetaE}f-g). 
The advective response $\Delta(-u\frac{\partial \theta_e}{\partial r})$ also contributes to this initial reduction in $\theta_e$ in the first couple of hours, during which the advective import of low $\theta_e$ air is reduced very near the surface at the base of the eyewall ($r=30-50 \; km$) but enhanced above this region and at smaller radii just inside of the original eyewall due to the frictionally-enhanced overturning circulation (Supplementary Figure 3). Similar to the drying experiment, the convective instability of boundary layer parcels near the original eyewall is reduced (Fig.\ref{thetaE}j). This effect thermodynamically suppresses the deep overturning through the free troposphere (Fig.\ref{thetaE}f-g). Initially, though, this thermodynamic suppression is smaller than the dynamic enhancement induced by surface roughening for $\tau\le1\;h$ below $z=4\;km$. The competing effect is evident at $\tau = 1\;h$ (Fig.\ref{thetaE}g), where $\Delta \vec{U}$ is positive below $z=4\;km$ and negative above 4-km height. Moreover, the response of outward flow at $z=4\;km$ feeds into a new region of upward motion that begins to emerge outside of the original eyewall at this time. Meanwhile, by $\tau=6\;h$, $\theta_e$ is strongly reduced near the surface beneath the original eyewall (Fig.\ref{thetaE}h), and thus the original eyewall is strongly stabilized. Therefore, the response of the overturning circulation outside the original eyewall is to shift radially-outward to where mid-level $\theta_e$ values are lower, and hence will be more unstable to boundary-layer parcels and thus more conducive to upward motion  (Fig.\ref{thetaE}a and g-h). However, for surface roughening, the initial Ekman enhancement of the overturning circulation also has a component that extends inside the original eyewall, locally enhancing the updrafts there (Fig.\ref{cir}f-g). The radius of maximum wind ultimately shifts consistently inwards (Fig.\ref{llwindfields}c), apparently following the inner updraft. With a weakened overturning circulation, the advective response $\Delta(-u\frac{\partial \theta_e}{\partial r})$ becomes generally positive (Supplementary Figure 3), indicating that the reduction in $\theta_e$ is driven by the reduction in surface fluxes rather than the advective response, similar to the drying experiment. Finally, boundary-layer $\theta_e$ continues to decrease and thus become increasingly stable for $\tau=24\;h$ (Fig.\ref{thetaE}i), thereby gradually weakening the deep overturning circulation through the depth of the troposphere within the inner core of the vortex (Fig.\ref{cir}i).
\begin{figure*}[t]
\centerline{\includegraphics[width=0.9\textwidth]{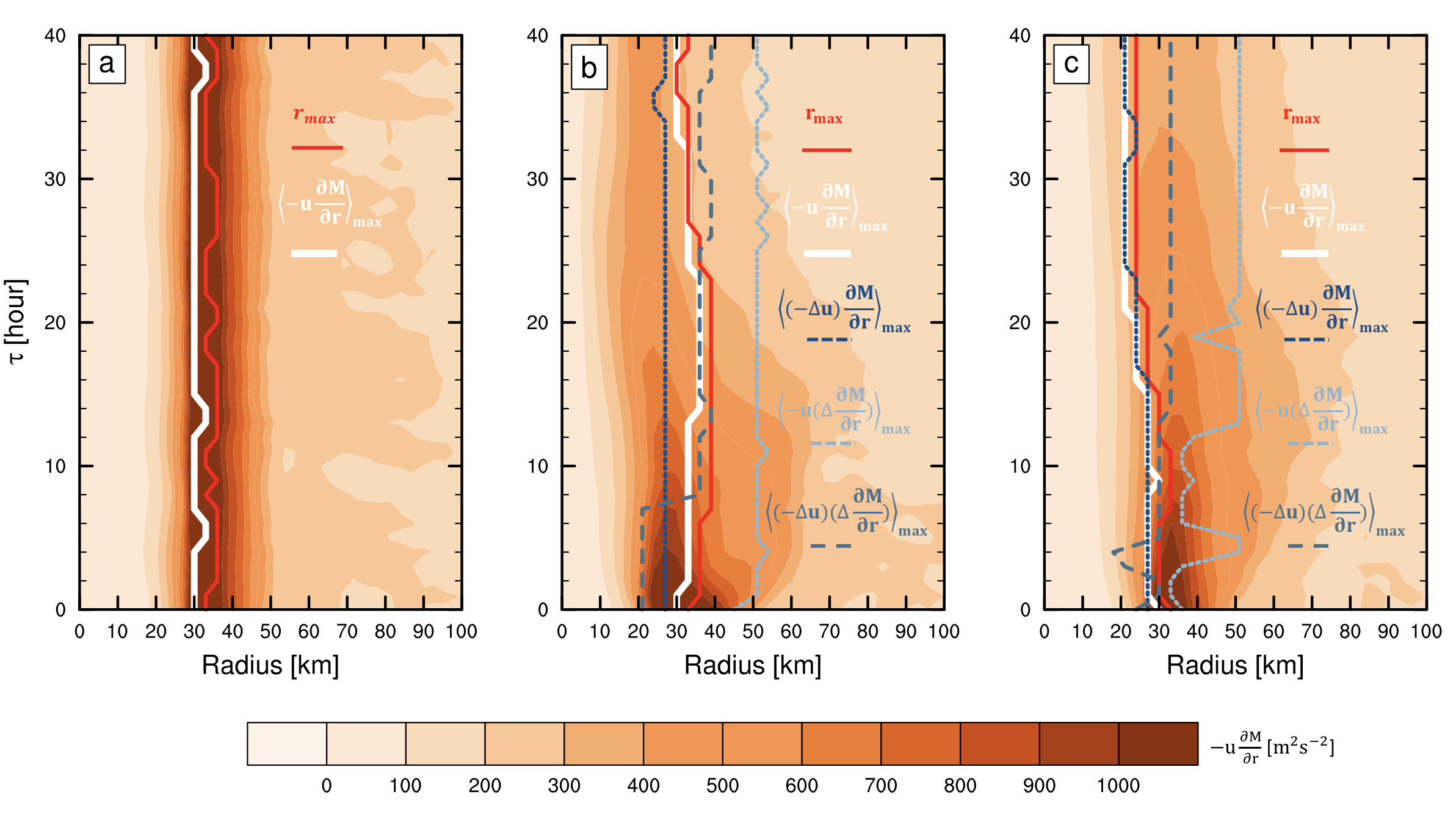}}
\caption{Hovm\"{o}ller diagram of $-u\frac{\partial M}{\partial r}$ on the lowest model level (color), with time-series of $r_{max}$ (red line) and the radius of maximum $-u\frac{\partial M}{\partial r}$ (white line) for a) CTRL, b) $0.25\epsilon$, and c) $4C_d$ from $\tau=0$ to $40\;h$. Also shown in (b) and (c): the radii of maximum $(-\Delta u)\frac{\partial M}{\partial r}$, $-u(\Delta\frac{\partial M}{\partial r})$ and the residual (see legend).}\label{rmaxanalysis}
\end{figure*}

Overall, while both surface forcings eventually weaken the storm to a similar state, their respective paths to re-equilibration are markedly different, particularly during the first 24 hours. First, the response of the overturning circulation is directly impacted by decreased surface moisture fluxes, through which the eyewall updraft can be stabilized. For surface drying, this arises due to the direct reduction of $F_{qv}$. For surface roughening, $F_{qv}$ is reduced by the rapidly weakened near-surface wind field, a process that is temporarily amplified by the enhanced advective import of low $\theta_e$ air by the Ekman-enhanced radial inflow. This additional, short-term frictional enhancement of the overturning circulation forced by surface roughening can temporarily overcome the effect of eyewall stabilization within the lower troposphere to continue to produce ascent. This secondary effect also enhances the import of angular momentum, though not enough to fully offset the enhanced loss of angular momentum by surface friction. Thus, the transient response to surface roughening is especially complex, as the primary circulation is directly suppressed dynamically and the secondary circulation is thermodynamically suppressed but dynamically enhanced.

\subsubsection{Evolution of $r_{max}$}

One particularly thorny yet important quantity is the radius of maximum wind, which lacks a simple governing equation to explain its dynamics. Here we apply our angular momentum response budget framework, which may provide some useful insight into this problem.

As noted in the previous subsection, variability in $r_{max}$ ought to be intimately linked to horizontal advection of angular momentum, $-u\frac{\partial M}{\partial r}$, which depends on the combined responses of the inflow and the vortex structure. The horizontal advection tendency $-u\frac{\partial M}{\partial r}$ at the lowest model level is shown in Fig.\ref{rmaxanalysis} across CTRL, $0.25\epsilon$, and $4C_d$. Notably, the location of maximum $\langle-u\frac{\partial M}{\partial r}\rangle_{max}$, lies just inside of $r_{max}$ and its temporal variability consistently follows $r_{max}$ across all three experiments.  Thus, $\langle -u\frac{\partial M}{\partial r}\rangle_{max}$ may offer an avenue for understanding the evolution of $r_{max}$ using the response budget decomposition given by Eq. \eqref{ADVanalysis}. Specifically, for each term in each term in Eq. \eqref{ADVanalysis}, we quantify how the location of its maximum value evolves in time (denoted with $\langle\rangle_{max}$). The results are displayed in Fig.\ref{rmaxanalysis}. We emphasize that this should not be interpreted as a true budget; it is simply a diagnostic tool for comparing responses due to the overturning circulation and the vortex structure. For surface drying, $\langle-u\frac{\partial M}{\partial r}\rangle_{max}$ lies between $\langle(-\Delta u)\frac{\partial M}{\partial r}\rangle_{max}$ and $\langle-u(\Delta\frac{\partial M}{\partial r})\rangle_{max}$ (Fig.\ref{rmaxanalysis}b), indicating that neither the overturning response nor vortex structure response is dominant in setting $r_{max}$. For surface roughening, though, $\langle(-\Delta u)\frac{\partial M}{\partial r}\rangle_{max}$ closely follows $\langle-u\frac{\partial M}{\partial r}\rangle_{max}$, suggesting a dominant role of the overturning response in setting $r_{max}$ (Fig.\ref{rmaxanalysis}c). The above analysis may be valuable for deeper analysis of changes in $r_{max}$ in future work.

\subsection{Precipitation field}\label{precipitation}

In addition to the low-level wind field, the precipitation response is also intimately dependent on the response of the secondary circulation in each experiment. To further analyze this relationship, we decompose the response of total precipitation (Fig.\ref{prcpall}) into dynamic and thermodynamic components using a simple precipitation scaling, $\widetilde{P}$ \citep{Emori+Brown2005}, given by
\begin{equation}\label{Eq_precip}
\widetilde{P}\sim \widetilde{w}\cdot \widetilde{q}
\end{equation}.

\begin{figure}[t]
\centerline{\includegraphics[width=0.5\textwidth]{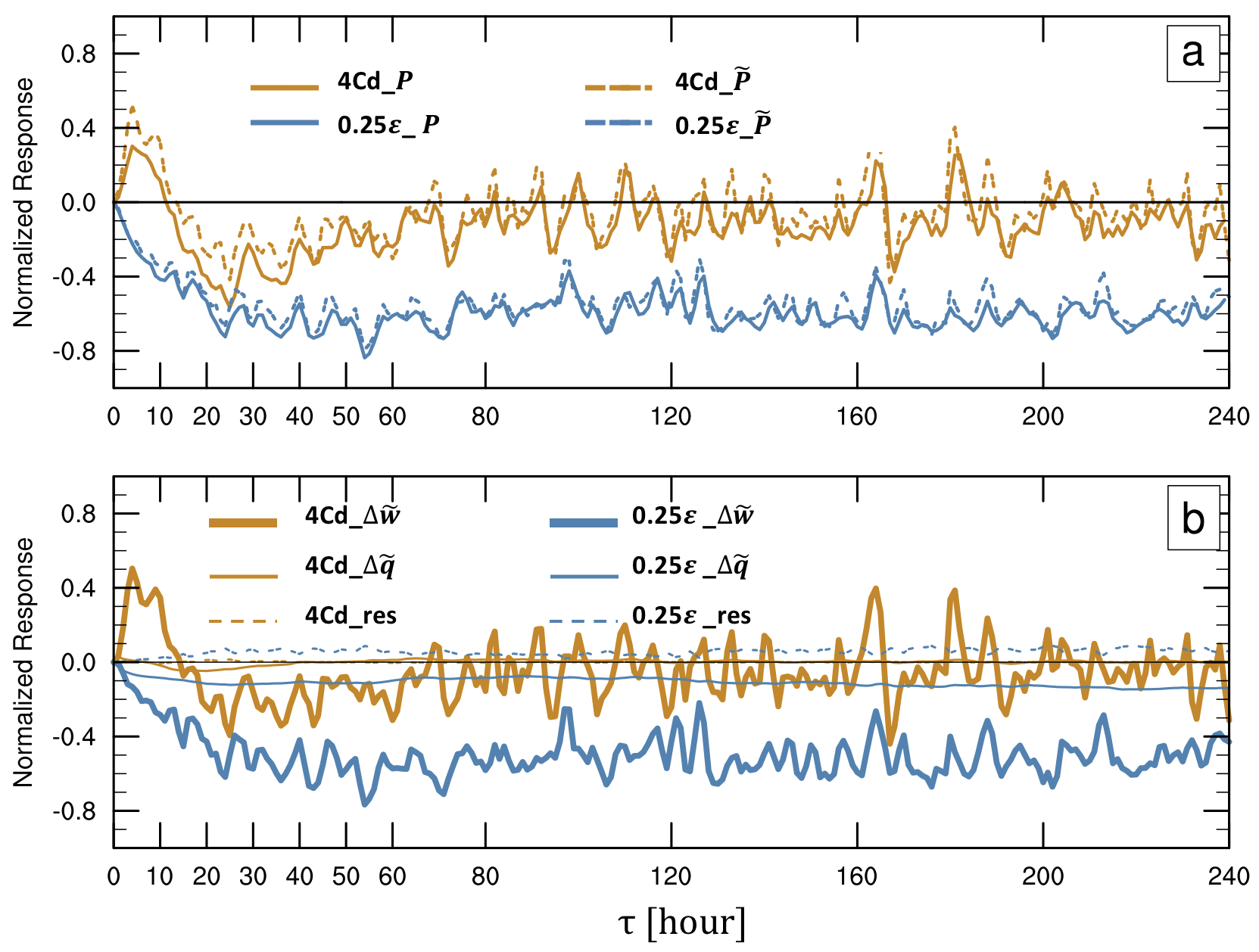}}
\caption{ (a) Temporal evolution of normalized responses in total precipitation $\frac{\triangle{P}}{P_{CTRL}}$ and simple decomposition precipitation $\frac{\triangle{\widetilde{P}}}{\widetilde{P}_{CTRL}}$ (Eq.\ref{precip_decomp}) for $4C_d$ and $0.25\epsilon$. (b) Temporal evolution of $\triangle{\widetilde{P}}$ normalized decomposition, including dynamic component $\frac{\triangle{\widetilde{w}}}{\widetilde{w}_{CTRL}}$, thermodynamic component $\frac{\triangle{\widetilde{q}}}{\widetilde{q}_{CTRL}}$ and non-linear residual (Eq.\ref{precip_decomp_diff}). The total precipitation $P$ is defined as the sum of the hourly precipitation within $r=0-150\;km$. Here $\widetilde{w}$ is the mean vertical velocity between $r=0-150\;km, z=1-4\;km$, representing the dynamic component; $\widetilde{q}$ is the mean 2-m mixing ratio within $r=0-150\;km$ representing the thermodynamic component.}\label{comP}
\end{figure}

The dynamical component, $\widetilde{w}$, is defined as the mean vertical velocity in the volume bounded by $z=1-4 \;km$, $r=0-150 \; km$, and the thermodynamic component, $\widetilde{q}$, is defined as the mean 2-meter mixing ratio within $r=0-150 \; km$. We may then use Eq. \eqref{Eq_precip} to linearly decompose changes in precipitation \citep{Bony+Emanuel2001}, i.e.
\begin{equation}\label{precip_decomp}
\triangle{\widetilde{P}}\approx(\triangle{\widetilde{w}})\cdot \widetilde{q}+\widetilde{w}\cdot(\triangle{\widetilde{q}})+(\triangle{\widetilde{w}})\cdot(\triangle{\widetilde{q}})
\end{equation}
\begin{equation}\label{precip_decomp_diff}
\frac{\triangle{\widetilde{P}}}{\widetilde{P}_{CTRL}}\approx\frac{\triangle{\widetilde{w}}}{\widetilde{w}_{CTRL}}+\frac{\triangle{\widetilde{q}}}{\widetilde{q}_{CTRL}}+\frac{(\triangle{\widetilde{w}})\cdot(\triangle{\widetilde{q}})}{\widetilde{P}_{CTRL}}
\end{equation}
where the first term on the RHS of Eq.\ref{precip_decomp_diff} is the normalized linear response of $\widetilde{P}$ due to changes in dynamic component $\widetilde{w}$, the second term on the RHS represents the normalized linear response of $\widetilde{P}$ due to changes in thermodynamic component $\widetilde{q}$, and the final term is the non-linear residual. 
We first compare the normalized responses of model-simulated total precipitation $\frac{\triangle{P}}{P_{CTRL}}$ and our simplified precipitation variable $\frac{\triangle{\widetilde{P}}}{\widetilde{P}_{CTRL}}$ for each experiment (Fig.\ref{comP}a). Indeed, the normalized responses of $\widetilde{P}$ closely follows the evolution of $P$.

For surface drying, the reduction in surface fluxes directly weakens the thermodynamic component, with a reduction of 10\% that is relatively constant out to 60 hours. Meanwhile, the stabilization of the eyewall updraft causes a significantly stronger dynamic suppression of precipitation, which increases in magnitude quasi-exponentially out to $\tau=40\;h$ (Fig.\ref{comP}b). This dynamic weakening dominates the decrease in total precipitation throughout the experiment.

For surface roughening, the dynamic response ($\triangle{\widetilde{w}}$) is the dominant control of variability in $\triangle{\widetilde{P}}$ throughout the evolution (Fig.\ref{comP}b), reaching a maximum enhancement of +50\% relative to CTRL at $\tau=5 \; h$ and then a maximum reduction of -40\% relative to CTRL at $\tau=24h$. In contrast, the thermodynamic response ($\triangle{\widetilde{q}}$) remains small and relatively constant throughout the evolution, with only a modest decrease over the first 36 hours. This result aligns with the earlier findings that there is an enhancement of the overturning circulation during the first 12 hours for roughening. Thus, the total precipitation remains higher than CTRL in $4C_d$ experiment until $\tau=10\;h$. Thereafter, the overturning circulation decays as explained in previous section, which weakens the eyewall updraft and directly reduces precipitation.

\section{Conclusions}\label{summary}

Landfalling tropical cyclones produce significant inland hazards, particularly high winds and rainfall-induced flooding. These hazards are intrinsically tied to the response of the TC low-level wind field after landfall. Two key surface forcings associated with the transition from ocean to land are drying and roughening. Here we have tested the response of the wind and precipitation fields in idealized axisymmetric numerical model experiments in which the surface drag coefficient ($C_d$) or the surface evaporative fraction ($\epsilon$) is instantaneously modified over a range of values of each to roughen or dry the surface beneath a mature storm. We characterized the temporal responses of TC intensity, size, and precipitation across experiments and then quantified the physical processes that underlie these responses, focusing on one representative experiment from each set that carries similar predicted intensity responses based on potential intensity theory. 
\begin{figure*}[t]
\centerline{\includegraphics[width=\textwidth]{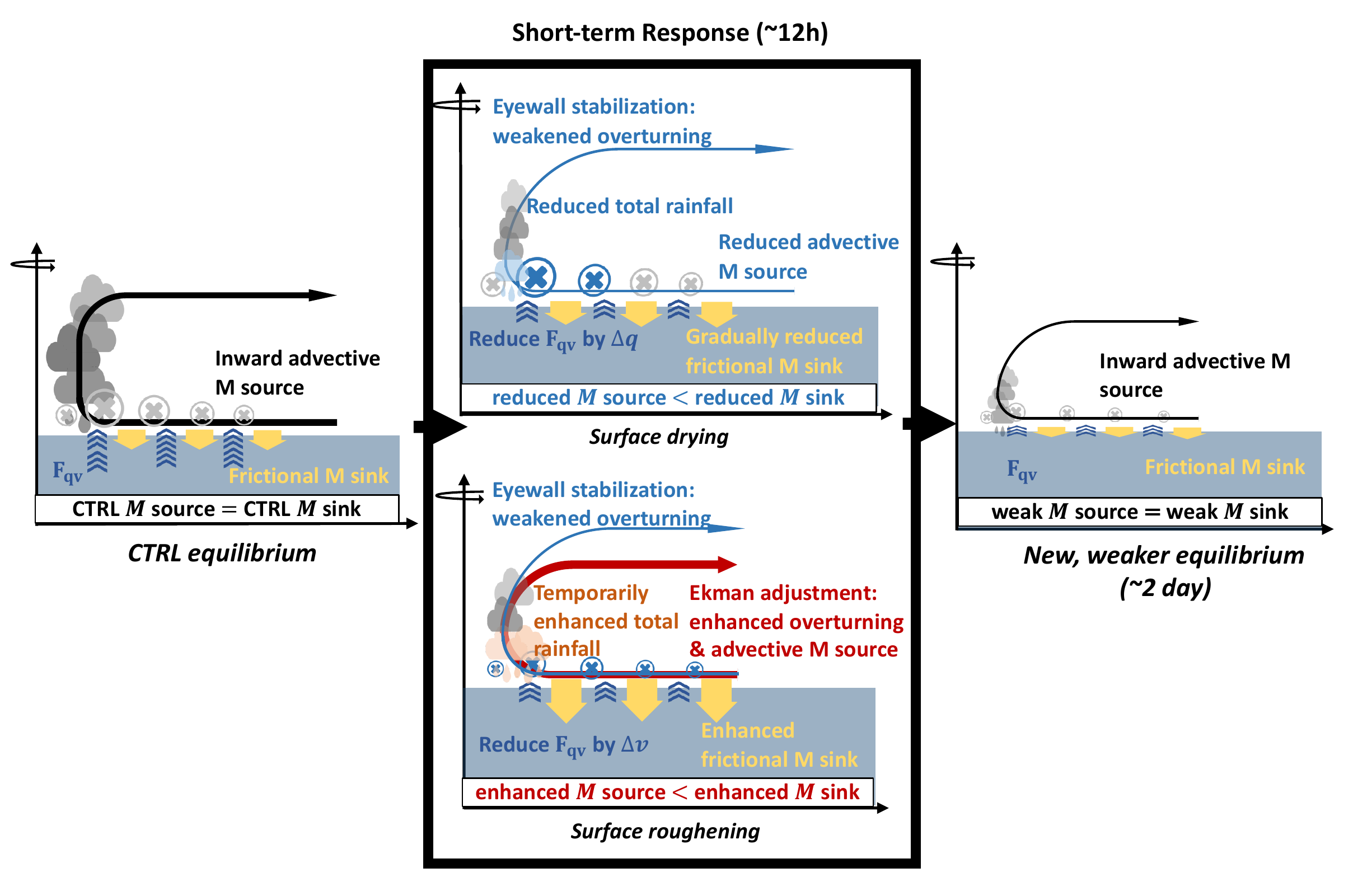}}
\caption{A simple conceptual schematic of the short-term response and long-term equilibrium to surface roughening or drying, including the primary circulation (crossed circle), overturning circulation and inward $M$ advective source, surface moisture fluxes $F_{qv}$, eyewall and precipitation, frictional $M$ sink. Response magnitude of each quantity is indicated by the icon width or size while the enhancement or reduction is colored in red or blue, respectively. Physical mechanisms for each response, including key parameters, are colored in the same manner.}\label{cartoon2}
\end{figure*}

Key findings are as follows:

\begin{enumerate}

\item Surface drying

Surface drying induces a single dominant response whose time-scale scales with the magnitude of drying. The primary circulation gradually weakens only within the inner-core in response to the gradual eyewall stabilization due to the rapid reduction in surface moisture fluxes. This stabilization progressively weakens the overturning circulation and hence reduces the advective source of angular momentum near the eyewall. This dynamical response in the overturning circulation is the dominant factor in the suppression of precipitation, with additional smaller reductions due to reduced boundary layer moisture. The response of the low-level wind field is initially strongest near $r_{max}$ and decreases with radius, such that the outer circulation remains relatively constant; $r_{34kt}$ decreases slowly with the weakening storm. The $r_{max}$ response is non-monotonic, as it initially increases for weak to moderate drying but decreases with strong drying; all cases eventually re-equilibrate to a value comparable slightly smaller than CTRL.

\item Surface roughening

Surface roughening induces responses on two distinct time-scales. The primary circulation is nearly instantaneously weakened at all radii near the surface due to the direct enhancement of the angular momentum sink due to surface friction. Meanwhile, the secondary circulation is temporarily enhanced due to the strengthened frictionally-induced inflow, despite the thermodynamic eyewall stabilization caused by both reduced moisture fluxes due to the reduced near-surface wind speeds and enhanced advection of low-$\theta_e$ air. This strengthened overturning circulation also temporarily increases precipitation within the eyewall region while eventually spinning down the vortex aloft by importing low angular momentum fluid out of the boundary layer. The overturning circulation subsequently weakens after 12 hours, as eyewall stabilization gradually weakens the entire overturning circulation similar to surface drying. These responses also act to reduce precipitation back towards the Control value. Storm size decreases monotonically at all radii as the storm weakens. 

\end{enumerate}

The above findings are summarized schematically in Figure \ref{cartoon2}. For surface drying, the feedback between the vortex and surface heat fluxes is weakened directly by reducing $F_{qv}$, which weakens the vortex slowly by gradually weakening the convective overturning circulation via stabilization of the eyewall that suppresses ascent. Thus, the overturning response exhibits a single time scale throughout the evolution. For surface roughening, the feedback is diminished by reducing $F_{qv}$ due to the strongly-decreased near-surface wind speeds of the vortex associated with enhanced surface friction, which also weakens the convective overturning circulation and further decays the vortex. However, this overturning response emerges following an initial period of frictional enhancement, resulting in a two-stage response to surface roughening. Though decaying through different pathways, the long-term transient responses to roughening and drying are similar, leaving a smaller and weaker vortex with a shallower circulation.  

Our findings indicate distinct responses of the wind and precipitation fields to surface roughening and drying on short time-scales ($\sim 0-2 \; days$) relevant to real-world landfall. This suggests that variations in surface land cover and moisture in the vicinity of storm landfall in the real world may have important effects on the associated inland hazards. Based on the results here, landfall over flat, smooth, dry land, may allow a storm to sustain its circulation for a longer duration despite the strong reduction in surface latent heat fluxes. In contrast, landfall over rough, moist land may rapidly weaken the storm wind field while greatly enhancing rainfall production. This work is a key first step towards understanding the physics of inland TC hazards, which is critical to understanding how inland hazards may change in a future climate state.

While we have attempted to characterize and understand the detailed responses to idealized surface roughening and drying, certain aspects were left for future work. First, the complex response of $r_{max}$ for different magnitudes of surface drying is not explained here. A deeper analysis of the long-term response evolution beyond $\tau=40\;h$ may be insightful to understand how the storm responds as the large-scale environment itself gradually evolves. It's unclear why our experiments with frictional enhancement did not produce a transient intensification period, in contrast to past work \citep{Montgomery2001}. Moreover, results may ultimately be sensitive to the details of boundary layer eddies at very fine-scales, which are necessarily parameterized in our model. Details of the boundary layer dynamics may require even higher resolution to reproduce the fine-scale structure and evolution of the low-level wind field.

Moving beyond this work, many open questions remain regarding the processes involved in landfall: How does $C_k$ and the ratio of $\frac{C_k}{C_d}$ vary over the land? How do antecedent soil moisture and storm precipitation affect surface moisture fluxes? How does the size and location of land relative to the storm center alter these responses? How will these axisymmetric responses change when adding real-world complexity, including three-dimensional geometry? More generally, the complex real world clearly includes a range of additional types of variability in the transition from ocean to land, including but not limited to sub-surface soil properties, ocean upwelling, coastline geometry, and topography, all of which may influence the evolution of low-level wind field and precipitation fields in real storms. In particular, we noted above that sensible heat fluxes may play a more significant role well inland at low intensities \citep{Kieu2015}. Sensible heat fluxes are sensitive to land surface properties and have large diurnal and spatial variation. In our work with fixed surface temperatures, sensible heat fluxes play a minimal role. However, idealized experiments with a coupled land-atmosphere model that allows for land feedbacks may yield different results.

The second part of this work will tackle two specific topics closely related to this work. First, what is the response to roughening if $C_k$ is allowed to scale with $C_d$? Second, how do existing theories for intensification, size, and structure compare against our idealized experiment sets? Both are natural extensions of the analyses presented here.

Otherwise, future work can test the responses in analogous three-dimensional modeling experiments to examine how our axisymmetric results are altered in the presence of resolved azimuthal vortex asymmetries. A finite storm translation speed can be applied to mimic the storm motion. Such three-dimensional experiments can serve as a bridge from our axisymmetric results towards a more realistic landfall while still maintaining sufficient simplicity to permit rigorous quantitative analysis and understanding. Finally, comparing these idealized studies with analyses of observational data and high-resolution numerical model outputs would help to link this simple understanding with the real-world and identify what types of complexities are most important for properly modeling inland hazards in nature.

\acknowledgments
The authors thank for all conversations related to this research during 22nd AMS conference on Atmospheric and Oceanic Fluid Dynamics and 19th Cyclone Workshop.


\appendix[A] 

We may quantify the relative contributions of surface sensible heat fluxes, $F_{SH}$, and latent heat fluxes, $F_{LH}$, to the potential intensity, $V_p$, for our CTRL simulation. We compare the baseline $V_p$ with its value,  calculated by assuming that the latent heat fluxes, sensible heat fluxes, or both are reduced by the reduction fraction $\epsilon$ ranging from 0 to 1. We denote the latter as ${V_p(\epsilon)}_X$, where $X$ indicates the flux type that is modified. For example, the relative importance of $F_{LH}$ is given by 
\begin{equation}\label{vp_lh}
   \frac{{V_p(\epsilon)}_{LH}}{V_p}=\frac{\sqrt{\frac{C_k}{C_d}\eta (F_{SH} + \epsilon F_{LH})}}{\sqrt{\frac{C_k}{C_d}\eta (F_{SH} +F_{LH})}}
\end{equation} 

\setcounter{figure}{0}
\begin{figure}[t]
\centerline{\includegraphics[width=20pc]{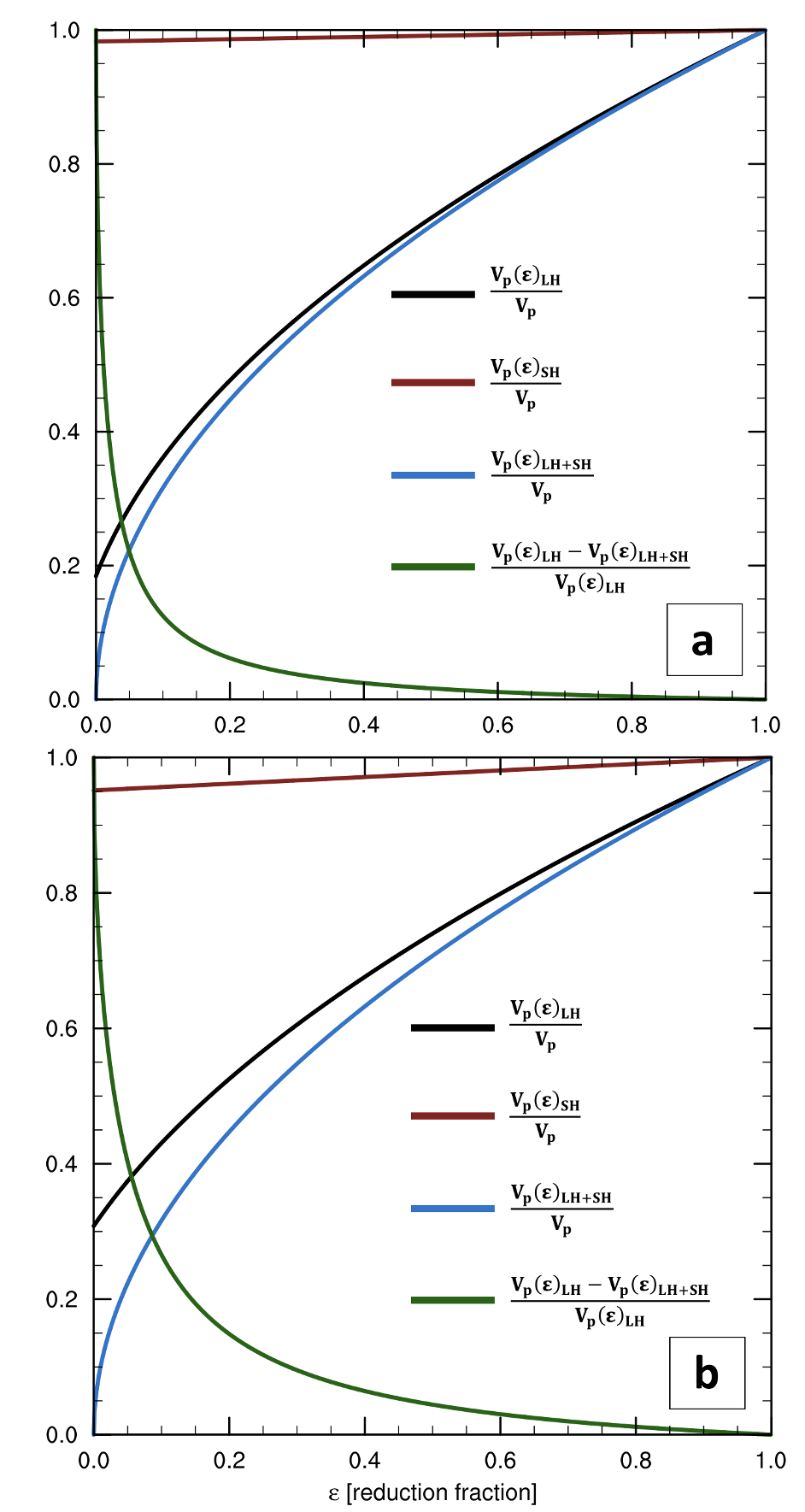}}
\appendcaption{A1}{Ratio of $\frac{{V_p(\epsilon)}_{X}}{V_p}$ with surface flux reduction fraction $\epsilon$ ranging from 0 to 1. $X$ denotes the type of surface flux that is modified (latent heat: $LH$ (black curve); sensible heat: $SH$ (red curve); or both (blue curve)). The ratio $\frac{{V_p(\epsilon)}_{LH}-{V_p(\epsilon)}_{LH+SH}}{{V_p(\epsilon)}_{LH}}$ is also shown (green curve). Baseline $V_p$ is calculated using Eq.\eqref{EQvp} from our Control using air-sea disequilibrium values from (a) the environment, and (b) at $r_{max}$.}\label{Vp_comp}
\end{figure}
Fig.A1a displays $V_p$ calculated using $\epsilon$ applied to just latent heat fluxes (black curve), just sensible heat fluxes (red curve), and both latent and sensible heat fluxes (blue), each as a fraction of the true $V_p$. Fig.A1a further displays the fractional contribution of the sensible heat fluxes to $V_p$ as $\epsilon$ is decreased (green curve), $\frac{{V_p(\epsilon)}_{LH}-{V_p(\epsilon)}_{LH+SH}}{{V_p(\epsilon)}_{LH}}$. This ratio indicates that the role of sensible heat fluxes remains small ($<10$\%) except when the latent heat fluxes are very strongly reduced to $\epsilon \le 0.1$. For this very strong drying, sensible heat fluxes begin to provide a significant fraction of the total surface heat fluxes and thus to the total $V_p$. Results are qualitatively similar when calculating $V_p$ using $F_{SH}$ and $F_{LH}$ at $r_{max}$ (Fig.A1b), though with a stronger contribution from sensible heat fluxes ($\epsilon<0.4$). This is because $\Delta T$ in Eq.\eqref{EQvp} is much higher at $r_{max}$ (0.832K) than the environmental $\Delta T$ (0.33K). The increasing importance of sensible heat fluxes with strong drying highlights that land surface properties, particularly near the eyewall, may become important for weak inland storms, as has been noted in past studies \citep{Kieu2015}.

 
\bibliographystyle{ametsoc2014}
\bibliography{references}

\end{document}